\documentclass[sts,preprint]{imsart}

\usepackage{amsthm,amsmath,natbib}
\RequirePackage[colorlinks,citecolor=black ,urlcolor=blue]{hyperref}

\startlocaldefs
\usepackage{macros_scharf}
\usepackage{amssymb, bbm, tabularx, multirow, amsmath, epsfig, threeparttable, amstext, subcaption}
\usepackage{graphicx}
\endlocaldefs

\begin{document}

\begin{frontmatter}

% "Title of the paper"
\title{Hierarchical approaches for flexible and interpretable binary regression models}
\runtitle{Hierarchical binary regression models}

% indicate corresponding author with \corref{}
\author{\fnms{Henry R.} \snm{Scharf}\corref{}
\ead[label=e1]{henry.scharf@colostate.edu}
% \thanksref{t1}
}
% \thankstext{t1}{Thanks to somebody}
\address{H. R. Scharf is Postdoctoral Fellow, Department of Statistics, Colorado State University \printead{e1}.}
% \affiliation{Colorado State University}
%
\author{\fnms{Xinyi} \snm{Lu}}
\address{X. Lu is Graduate Student, Department of Statistics, Colorado State University.}
% \affiliation{Colorado State University}
%
\author{\fnms{Perry J.} \snm{Williams}}
\address{P. J. Williams is Assistant Professor, Department of Natural Resources and Environmental Science, University of Nevada.}
% \affiliation{University of Nevada}
\and
\author{\fnms{Mevin B.} \snm{Hooten}}
\address{M. B. Hooten is Assistant Unit Leader, U.S. Geological Survey, Colorado Cooperative Fish and Wildlife Research Unit, and Professor, Departments of Fish, Wildlife, \& Conservation Biology and Statistics, Colorado State University.}
% \affiliation{U.S. Geological Survey}
% \affiliation{Colorado State University}

\runauthor{Henry R. Scharf}

\begin{abstract}
Binary regression models are ubiquitous in virtually every scientific field. Frequently, traditional generalized linear models fail to capture the variability in the probability surface that gives rise to the binary observations and novel methodology is required. This has generated a substantial literature comprised of binary regression models motivated by various applications. We describe a novel organization of generalizations to traditional binary regression methods based on the familiar three-part structure of generalized linear models (random component, systematic component, link function). This new perspective facilitates both the comparison of existing approaches, and the development of novel, flexible models with interpretable parameters that capture application-specific data generating mechanisms. We use our proposed organizational structure to discuss some concerns with certain existing models for binary data based on quantile regression. We then use the framework to develop several new binary regression models tailored to occupancy data for European red squirrels (\textit{Sciurus vulgaris}).
\end{abstract}

% \begin{keyword}[class=MSC]
% \kwd[Primary ]{Generalized linear models}
% \kwd{General nonlinear regression}
% \end{keyword}

\begin{keyword}
\kwd{Binary response}
\kwd{Generalized linear models}
\end{keyword}

\end{frontmatter}

\section{Introduction}

In almost every area of scientific application, binary regression models are used to understand how the probability of a ``success'' depends on covariates of interest. In most cases, generalized linear models (GLMs) are employed to learn about this probability curve. While the traditional three-part structure of random component, systematic component, and link function provides researchers some flexibility in the parametric form of the probability curve, in practice these components are often specified based on commonly used defaults and without consideration of the real natural phenomena that generated the observed binary data. Moreover, the most widely used GLMs often require rigid assumptions that are unrealistic. The need for flexible binary regression models that relax assumptions about the data generating process has resulted in a substantial literature motivated by various applications. In this review, we describe a novel organization of generalizations of traditional binary regression methods based on the familiar three-part structure of GLMs. We use our proposed organizational structure to discuss some concerns with certain existing models for binary data based on quantile regression. Then, we use the framework to develop several new binary regression models tailored to occupancy data for European red squirrels (\textit{Sciurus vulgaris}), and score each model based on its predictive ability. The conclusions from this exploration highlight the need for researchers to consider flexibility, interpretability, and predictive power when selecting from several candidate models.

\subsection{Background and notation}
Denote by $\text{Pr}(y_i = 1)$ the probability that a binary response random variable $y_i$ is 1 (i.e., that we observe a ``success'' for the $i\text{th}$ datum), where $i = 1, \dots, n$ indexes the observations. The goal of a binary regression analysis is to infer the relationship between these probabilities and a vector of relevant covariates, $\mathbf{x}_i$, that takes on values from the covariate space $\mathcal{X}$  (we include the subscript, $i$, when it is helpful to emphasize the association between a particular fixed vector of covariate values and the response, $y_i$). Thus, interest lies in learning about the characteristics of a map $p: \mathcal{X} \rightarrow \left[ 0, 1 \right]$, parameterized by a (possibly infinitely long) vector, $\boldsymbol\theta$. The data are generally assumed to be conditionally independent given the covariates, leading to a log-likelihood of the form

\begin{align}
  l(\boldsymbol\theta | \mathbf{y}) = \sum_{i = 1}^n y_i \log(p_i) + (1 - y_i) \log(1 - p_i),
    \label{eqn:likelihood}
\end{align}

where $\mathbf{y} = \lp y_1 \dots y_n \rp'$, $\mathbf{X} = \lp \mathbf{x}_1 \dots \mathbf{x}_n \rp$ and $p_i = p(\mathbf{x}_i, \boldsymbol\theta)$. Likelihood-based approaches may be used to fit the model to data. Several widely-used software packages for both frequentist and Bayesian paradigms are available to facilitate model fitting.

By far the most common framework employed to specify a functional form for the probability curve is a GLM given by

\begin{align}
  y_i &\sim f_Y(\mathbf{x}_i), \label{eqn:glm1}\\
  \eta(\mathbf{x}) &= \mathbf{x}' \boldsymbol\beta, \label{eqn:glm2}\\
  g(\E{y_i }) &= \eta(\mathbf{x}_i). \label{eqn:glm3}
\end{align}

GLMs are often described as consisting of these three parts. The first part, \eqref{eqn:glm1}, called the ``random component,'' specifies the probability density, $f_Y$, of the observed data. In the case of a binary response, the probability distribution must be Bernoulli with probability $p_i$. The second part, called the ``systematic component,'' defines a function $\eta(\mathbf{x})$ in \eqref{eqn:glm2} related to the expected value of the data. This function is a linear combination of covariates of interest with effects parameterized by $\boldsymbol\beta$. The third part, called the ``link function,'' is represented by the invertible function, $g(\cdot)$, in \eqref{eqn:glm3}. The purpose of the link function is to align the systematic component with the support of the expected value of the response distribution. For the case of binary regression, $\E{y_i} = p_i \in [0, 1]$, so the link must transform the unit interval to the full range of the systematic component, typically taken to be the whole real line. The most common form for the link in binary regression is the logit function, which, in tandem with a linear systematic component, yields the probability function $p(\mathbf{x}_i, \boldsymbol\beta) = \text{Pr}(y_i = 1) = g^{-1}(\mathbf{x}_i'\boldsymbol\beta) = 1 / (1 + e^{-\mathbf{x}_i'\boldsymbol\beta})$. GLMs are used for a wide range of response types; however, the remainder of the manuscript considers only binary data, and thus $f_Y$ is always taken to be Bernoulli. For a more in-depth treatment of GLMs, see for example \cite{McCullagh1983} and \cite{Agresti2002}.

In many applications, the traditional linear systematic component paired with a symmetric link function such as the logit is too rigid a form to adequately explain the variation in probability of an observed binary response. In these situations, traditional approaches must be generalized to allow for more flexible probability curves. There has been a great deal of focus on developing new methods that can account for variation in binary data showing evidence of residual heterogeneity beyond what is explainable with simple linear effects and a logit or probit (i.e., Gaussian CDF) link \citep[e.g.,][]{Bazan2010}. Methods proposed in the literature have come from a variety of different areas, with applications in psychology \citep[e.g.,][]{Fahrmeir2007}, ecology \citep[e.g.,][]{Augustin1996, Komori2016}, economics \citep[e.g.,][]{Khan2013}, and many other disciplines. We contend that generalizations of traditional GLMs can be usefully grouped based on which of the three GLM components are modified to allow for additional flexibility in the probability curve.

First, increased flexibility in the functional relationship between the probability of success and the covariates may be achieved through a relaxation of the linear form $\eta(\mathbf{x}) = \mathbf{x}'\boldsymbol\beta$ in the systematic component. For instance, a polynomial relationship (e.g., quadratic) is often considered for the $j\text{th}$ covariate when it is hypothesized that a ``peak'' effect exists at some unknown level. So-called non-parametric approaches attempt to relax assumptions about the probability curve by allowing $\eta(\mathbf{x})$ to be an arbitrary function of covariates \citep{Hastie1986, Hefley2017, Wood2017}. To ensure parameter identifiability, $\eta(\mathbf{x})$ is sometimes constrained to possess certain types of smoothness \citep[e.g.,][]{Wood2017} or monotonicity \citep[e.g.,][]{Meyer2008}.

Second, increased flexibility can arise through alternative specifications of the link function. This approach generalizes traditional logistic/probit regression by changing the functional form of $g(\E{y_i})$ to include additional unknown parameters to be estimated from the data. While several approaches have been proposed in the literature for binary regression models accommodating so-called asymmetric \citep{Chen1999, Komori2016, Maalouf2011} or heavy-tailed link functions \citep{Wang2010}, a clear motivation for specifying a model with a non-standard link function comes when the researcher possess knowledge about the application-specific data generation process.

Finally, flexibility in the probability curve may be introduced by modifying the random component. For binary data, the marginal distribution of the responses must be Bernoulli. However, dropping the assumption of conditional independence allows researchers to account for residual dependence in the observations while controlling for covariate effects \citep[e.g.,][]{Augustin1996}. One important reason residual dependence may be present in the data is the existence of important but unaccounted for covariates. Models for dependent binary data can be achieved through the introduction of random effects in the systematic component. Frequently, random effects are indexed by known or unobserved class structures in the data, or spatio-temporal information \citep[e.g.,][]{Goldenberg2010, Diggle1998}.

A single model may incorporate changes to more than one of these three components. Crucially, sources of flexibility in the probability curve introduced through multiple components do not act independently. For example, two different link functions, $g_a$ and $g_b$, paired with two potentially non-linear systematic components $\eta_a(\mathbf x)$ and $\eta_b(\mathbf x)$ produce exactly the same probability curve if

\begin{align}
  g_b^{-1} \left(\eta_b(\mathbf x)\right) = g^{-1}_a\left(\eta_a(\mathbf x)\right), \label{eqn:equiv}
\end{align}

since this implies that the probability of success $p_i = \E{y_i} = g^{-1} \left(\eta(\mathbf{x}_i))\right)$ is the same under each pair of links and systematic functions. A corollary of this result is that a non-linear systematic component paired with one type of link function has an equivalent representation using a linear systematic component with another link function. Thus, the specification of the systematic component and the link function must be made holistically. Failing to consider the joint implications of choices made about each facet of a GLM can lead researchers to unintentionally specify models with parameters that are not identifiable from binary response data.

To facilitate comparison of the broad range of approaches in the literature and to aid practitioners interested in developing bespoke models for specific applications, we describe a hierarchical formulation for binary regression in Section~\ref{sec:auxiliary_var} that makes use of auxiliary variables. Representing models through auxiliary variables is particularly illuminating when deciding what link function is most appropriate. In some cases, the auxiliary variables may correspond to interpretable features of a hypothesized data generation process such that the resulting model has the benefit of being more realistic than traditional approaches with more interpretable parameters.

In Section~\ref{sec:mod_link}, we discuss generalizations of traditional link functions and how carefully choosing a link function can lead to more realistic models with increased interpretability. In Section~\ref{sec:mod_random}, we discuss the introduction of random effects and how this can be viewed as a relaxation of the traditional assumption of conditional independence in the random component. We discuss the motivations, benefits, and drawbacks of modifications to each of three GLM components through an application to ecological data in Section~\ref{sec:application}.

% Summarizing sentence: If you haven't hear of binary regression before, here's what it is, and here are the three ways GLMs may be generalized to account for residual heterogeneity.

\subsection{Auxiliary variable construction}\label{sec:auxiliary_var}

Statistical models for binary data are sometimes specified using an auxiliary variable construction such that

\begin{align}
\begin{split}
  y_i | z_i &= \mathbbm{1}_{z_i > 0}, \\
  z_i | \boldsymbol\beta &\sim f_Z (\mathbf{x}_i, \boldsymbol\beta),
\end{split} \label{eqn:auxiliary}
\end{align}

$\mathbbm{1}_{z_i > 0}$ is an indicator function that is 1 when $z_i$ is positive and 0 otherwise, and the conditional distribution $f_Z$ is a member of a family of probability density functions (PDFs) with parameters that depend on $\mathbf{x}_i$ and $\boldsymbol\beta$. The probability distribution for $y_i| \boldsymbol\beta$ is Bernoulli with probability of a success equal to the probability that the auxiliary variable, $z_i$, exceeds $0$, or $1 - F_Z(0| \boldsymbol\beta)$, where $F_Z$ is the cumulative distribution function for $f_Z$. An early well-known example of such a hierarchical construction came from \cite{Albert1993} who noted that probit regression arises when $f_Z$ is chosen to be a Gaussian density with mean $\E{z_i | \boldsymbol\beta} = \mathbf{x}_i'\boldsymbol\beta$ and variance 1. This is due to the symmetry of the Gaussian distribution which yields $1 - \Phi(-\mathbf{x}'\boldsymbol\beta) = \Phi(\mathbf{x}'\boldsymbol\beta)$. The benefit of the hierarchical specification in this case is that a conjugate prior exists for $\boldsymbol\beta$ (multivariate Gaussian) and model fitting can proceed from a Bayesian perspective using a Markov chain Monte Carlo (MCMC) algorithm comprised entirely of Gibbs updates, obviating the need to adjust tuning parameters required by other methods such as Metropolis-Hastings random walks. The approach has been used in a wide variety of applications including species occupancy models in ecology \citep[e.g.,][]{Hooten2003, Dorazio2012, Johnson2013}.

Hierarchical representations of binary regression models share a close connection with GLMs. In general, a hierarchically-specified binary regression model is equivalent to a GLM if there exists a link function and systematic component such that $g^{-1}\lp\eta(\mathbf{x}_i)\rp = 1 - F_Z(z_i = 0 | \boldsymbol\beta)$. When, as is commonly the case, the expected value of the auxiliary variable depends linearly on the covariates (i.e., $\E{z_i | \boldsymbol\beta} = \mathbf{x}_i'\boldsymbol\beta$) and $f_Z$ is a member of a location family of probability distributions, the model is guaranteed to be a GLM with systematic component, $\eta(\mathbf{x}_i) = \mathbf{x}_i'\boldsymbol\beta$, link function $g(\E{y_i}) = -F_Z^{-1}(1 - \E{y_i})$ (or equivalently, inverse link function $g^{-1}(\eta(\mathbf{x}_i)) = 1 - F_Z(-\eta(\mathbf{x}_i))$.

The flexibility afforded by these mild conditions allows for the development of binary regression models motivated by the unobserved natural phenomena that give rise to observed binary data. For example, consider an application in ecology where the data are site-specific observations of the presence ($y_i = 1$) or absence ($y_i = 0$) of some species of interest, and the covariates are features of the landscape measured at each site (e.g., vegetation cover, elevation, etc.). If we assume that observations are made with perfect detection (i.e., if a species is present in any abundance at site $i$, $p_i = 1$, and similarly, $p_i = 0$ when the species is absent), then we could define presence as an indicator that the unobserved abundance of the species, $z_i$, is greater than 0. A natural choice for $f_Z$ would then be a family of discrete, non-negative probability mass functions such as the Poisson distribution with rate parameter $\lambda = e^{\mathbf{x}'\boldsymbol\beta}$ \citep[as in][]{Royle2003}.

For the case of Poisson-distributed auxiliary variables, the probability of success is $p_i = \text{Pr}(z_i > 0 | \boldsymbol\beta) = 1 - \exp\left\{-e^{\mathbf{x}_i'\boldsymbol\beta}\right\}$. This corresponds to the special case of a GLM with linear systematic component paired with a complimentary log-log (cloglog) link function. Alternatively, one might specify a negative binomial distribution for $z_i$, in which case the model construction presents a generalization of the typical GLM through the inclusion of a link function with an additional unknown parameter. The family of negative binomial probability distributions is characterized by two parameters (one helpful parameterization uses a mean and an over-dispersion parameter), of which the single-parameter Poisson and logistic families are two particular limiting cases.

Just as there exist infinite combinations of systematic components and link functions resulting in the same probability curve, so too are there multiple equivalent auxiliary variable specifications of hierarchical models for binary data. The relationship given by equation~\eqref{eqn:equiv} has an analogous representation in the auxiliary variable framework. Two different auxiliary variables, $z_i^{(a)}$ and $z_i^{(b)}$ with respective CDFs, $F_Z^{(a)}$ and $F_Z^{(b)}$, describe the same probability curve for $y_i$ if

\begin{align}
  F_Z^{(a)}(z_i^{(a)} = 0) = F_Z^{(b)}(z_i^{(b)} = 0),
  \; \forall \; \mathbf{x}_i \in \mathcal{X}. \label{eqn:equiv_aux}
\end{align}

We note for completeness that the indicator function in~\eqref{eqn:auxiliary} could, in principle, partition the support of $z_i$ into any two complimentary sets. However, the positive/non-positive indicator is the most common in the literature and we will only consider examples using this partition in the present work.

% Summarizing sentence: Auxiliary variables are another way we sometimes specify models for binary data; here are some examples, and some reasons why we might prefer one over another.

\subsection{Scientific interpretation}

% - covariate effects
% - auxiliary variables
% - probability curve and its derivatives

When performing a regression analysis for binary data, researchers may be motivated by several potential questions of interest. In almost every case, there will be interest in estimating the true probabilities of success for each observed combination of covariates, as well as predicting probabilities for new combinations. In addition, researchers will often be interested to know how the probability of success changes for small perturbations in the covariates. For these types of questions, the particular parameterization of $p$ is immaterial. That is, provided a sufficiently flexible model for the probability map has been specified, it does not matter, asymptotically, what particular distribution the researcher has chosen for the auxiliary variables. Equivalently, given a sufficiently flexible systematic component, it does not matter what the functional form of the link function is, as long as it is invertible and maps the support of $\E{y_i}$ to the real line. Thus, if the questions of scientific interest concern only the raw probability function, the primary modeling considerations will be that the algorithm for fitting the model to data is efficient and numerically stable, and that there are useful tools available for assessing goodness of fit \citep[e.g.,][]{Conn2018, Wright2019}.

In every analysis of binary data, both the inferred probability curve and its derivative with respect to the covariates are natural quantities to investigate and interpret. The probability curve (or, more generally, surface when multiple covariates are considered) provides predictions for the probabilities of successes for both observed and unobserved combinations of covariates. Interpreting the probability curve is a standard component to statistical analysis of binary data in virtually all applications. Another valuable curve that is under-utilized in analyses of binary data is the gradient of the probability curve, $\nabla_\mathbf{x} p(\mathbf{x})$. The gradient of the probability curve is analogous to the coefficients in a traditional linear regression problem in that it describes how the probability of success changes for small perturbations in the levels of the covariates. However, unlike in traditional linear regression, the probability curve is a non-linear function of the covariates, and thus the value of the gradient is not constant, but rather depends on the value of $\mathbf{x}$. One useful region of the gradient surface to consider is the vicinity of the mean of the observed covariates, although features of the gradient surface near the extremes of the observed covariate values may also be scientifically relevant. For example, in actuarial applications, extreme regions of the covariate space might correspond to particularly high or low risk individuals.

In addition, the parameters $\boldsymbol\theta$ themselves may offer additional opportunities for scientific learning. For example, the regression coefficients in a traditional logistic regression model represent the change in log-odds of the response for an increase of 1 unit in the associated covariate. As the example involving Poisson random variables in the previous section showed, the auxiliary variables may have a useful mechanistic interpretation corresponding to some unobserved natural process. Because there is no unique auxiliary variable specification for a given probability curve, researchers must choose from an infinite collection of equivalent hierarchical models for binary data. The most useful models are those that are parameterized in a way that yields useful interpretation and/or admits efficient algorithms for model fitting. For instance, in the occupancy example mentioned in Section~\ref{sec:auxiliary_var} that assumes Poisson-distributed auxiliary variables with mean $e^{\mathbf{x}'\boldsymbol\beta}$, $\beta_j$ can be interpreted as the linear effect on log-abundance of increasing $x_j$ by 1 unit.

\section{Modifying the link function}\label{sec:mod_link}

One way to introduce flexibility in the probability curve is to retain the linear relationship of the traditional GLM approach, but use a flexible family of link functions with free parameters to be estimated from the data. Flexible link functions with unknown parameters have been proposed in the literature as a way to accommodate residual heterogeneity in the data exhibiting various characteristics. One common concern about traditional link functions that has motivated some of this work concerns the potentially restrictive assumption of symmetry. So-called symmetric link functions make the implicit assumption that there exists a sub-space of $\mathcal{X}$ defined by $\mathbf{x}'\boldsymbol\beta = 0$ around which changes in covariate values, $\Delta \mathbf{x}$, result in changes to the probability curve of a magnitude that only depends on the length of the vector $\Delta \mathbf{x}$. In particular, changes in probability for shifts $\Delta \mathbf{x}$ and $-\Delta \mathbf{x}$ are the same in size, but in opposite directions (i.e., $p(\Delta\mathbf{x}) = 1 - p(-\Delta\mathbf{x})$). Inverse link functions defined through location family auxiliary variables for which $1 - F_Z(-z) = F_Z(z)$ are examples of symmetric link functions. Another way to state this symmetry relation is to say that the gradient of the probability curve, $\nabla_{\mathbf{x}} p(\mathbf{x})$, is a symmetric function about $\lbr \mathbf{x} : \mathbf{x}'\boldsymbol\beta = 0 \rbr$.

There are several useful ways to relax the assumption of symmetry through the link function. For example, \cite{Komori2016} defined a new inverse link function of the form

\begin{align}
  g^{-1}(\eta(\mathbf{x})) = \frac{\exp\lbr\eta(\mathbf{x})\rbr + \kappa}
    {1 + \exp\lbr\eta(\mathbf{x})\rbr + \kappa}
\end{align}

that modifies traditional logistic regression. As $\kappa > 0$ grows, an increasing amount of ``right'' skewness enters the link function, in that $1 - p(-\Delta\mathbf{x}) < p(\Delta\mathbf{x})$. The cloglog link function is an example of a ``left''-skewed link function (i.e., $1 - p(-\Delta\mathbf{x}) > p(\Delta\mathbf{x})$) where the degree of skewness is determined by the systematic component, rather than a free parameter. \cite{Prentice1976} proposed a two parameter model for skew link functions that allowed for skewness to occur in either direction, and was among the first such generalizations of traditional symmetric approaches. Plots (c) and (g) in Figure~\ref{fig:matching_auxiliary} show examples of left- and right-skewed probability curves, respectively. Active development of new methodology for skew link functions is ongoing \citep[e.g.,][]{Lemonte2018}.

Introducing additional flexibility through parameters in the link function can indeed result in better-fitting models, yielding probability curves and gradients thereof that more closely align with the true data-generating mechanism. However, the skewness parameter used in the inverse link function of \cite{Komori2016} can be difficult to interpret directly, and its introduction complicates the log-odds interpretation of the regression coefficients. In addition, the skewness parameter can be highly confounded with the intercept term, resulting in potentially unstable estimation algorithms. An alternative method to account for skewness is based on an auxiliary variable hierarchical model in which the distribution of $z_i | \boldsymbol\beta$ is asymmetric. As noted already, specifying Poisson-distributed auxiliary variables with conditional mean $e^{\mathbf{x}'\boldsymbol\beta}$ is equivalent to a GLM with a cloglog link, but in this case the skewness in the probability curve has an intuitive interpretation. It arises as a direct result of the data generating process, which consists of observing whether a count process is positive or zero. In addition, the regression coefficients can be interpreted as the linear effect of each covariate on log-abundance in ecological applications. When the application does not present a natural choice for the auxiliary variable distribution, phenomenological models that incorporate asymmetric probability curves can still be constructed using the hierarchical representation \citep[e.g.,][]{Chen1999, Xing2017}.

It is possible, in the pursuit of ultimate model flexibility, to take a non-parametric view of the conditional auxiliary variable distribution and allow $z_i | \boldsymbol\beta$ to arise from any arbitrary location-family of probability distributions with location $\mathbf{x}_i'\boldsymbol\beta$, which permits the same level of flexibility as the non-parametric mean approach of \cite{Choudhuri2007}. Care must be taken when introducing flexibility through the auxiliary variables so that the parameters $\boldsymbol\theta$ remain identifiable. For example, as we discuss in Section~\ref{sec:QR}, quantile regression approaches for binary data, which attempt to make minimal assumptions about the auxiliary variables conditional distribution \citep[e.g.,][]{Manski1985, Benoit2012, Padellini2018}, have model parameters that are only identifiable up to a multiplicative constant.

\subsection{Issues with parameter identifiability}\label{sec:identifiability}

As with any statistical model, it is necessary to ensure that parameters in a binary regression model are identifiable from the data. Non-identifiability can sometimes be difficult to anticipate in hierarchically-specified models, as we demonstrate in the following simple example. Indeed, as we discuss in Section~\ref{sec:QR}, non-identifiability has sometimes gone undetected in the literature, potentially leading to unsubstantiated scientific conclusions.

\begin{figure}
\centering
\includegraphics[width = \linewidth]{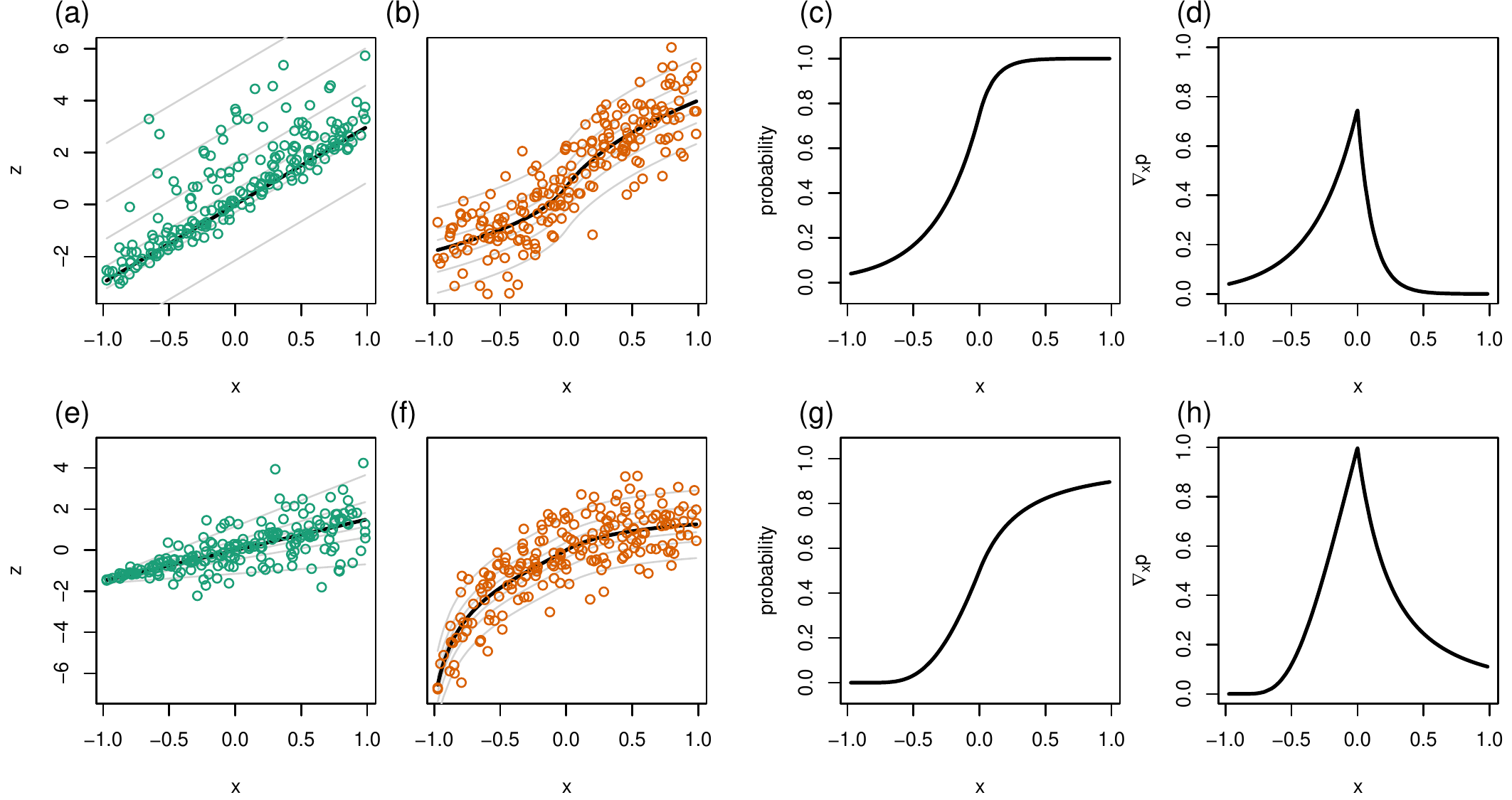}
\caption{Plots (a) and (b) show two different auxiliary variable distributions that result in the same left-skewed probability curve, (c), and probability gradient, (d). Plot (a) shows auxiliary variables that vary linearly with the covariate, $x$, and arise from an asymmetric Laplace distribution. Points represent realizations of the auxiliary variables and lines represent the mode and several quantiles of the auxiliary variables as a function of the covariate. Plot (b) shows normally-distributed auxiliary variables that have a non-linear relationship with the covariate. Plot (e) shows an example of heteroskedastic, normally-distributed auxiliary variables with means that vary linearly with the covariate. Plot (f) shows an equivalent representation of the right-skewed probability curve using normally-distributed, homoskedastic auxiliary variables with means that are a non-linear function of the covariate. Interpretation is analogous to the top row.}
\label{fig:matching_auxiliary}
\end{figure}

Specifying a model for binary data in a hierarchical form can provide intuition about the model and aid in comparing closely related models. However, it is important to note that many seemingly distinct statistical models based on auxiliary variables are equivalent in that they yield identical probabilities for the observed binary random variables and are therefore not identifiable from the data. For example, the top row of Figure~\ref{fig:matching_auxiliary} shows two different auxiliary variable specifications that satisfy~\eqref{eqn:equiv_aux}, and therefore produce equivalent conditional probabilities at the data level. In this example involving a single covariate, the auxiliary variables in (a) are asymmetrically distributed around a linearly varying trend, while the auxiliary variables in (b) are symmetrically distributed, but have a non-linear relationship with the covariate.

Plots (c) and (d) show the resulting equivalent probability curve and probability gradient. The bottom row of Figure~\ref{fig:matching_auxiliary} shows another pair of auxiliary variable specifications that result in the same right-skewed probability curve, with plots arranged analogously. Plot (e) shows an example of auxiliary variables that are normally-distributed and vary linearly and heteroskedastically with the covariate. Plot (f), as in (b), shows normally-distributed, homoskedastic auxiliary variables that have a non-linear relationship with the covariate.

Non-identifiability at the model level may appear to be fundamentally different than the familiar notion of parameter identifiability, since it is in a sense a statement about our ability to discern from the data which of two distinct models generated the data. To reconcile this false distinction, consider a subsuming model that permits both auxiliary variable specifications and includes a latent binary variable that indicates which auxiliary variable model is used to generate the data. To state that the two statistical models are not identifiable is equivalent to stating we cannot identify the latent binary-valued parameter.

All four model specifications shown in Figure~\ref{fig:matching_auxiliary} can also be represented using the three components of the GLM framework. In the top row, (a) corresponds to a binary regression model with a linear systematic component and asymmetric inverse link function defined by the CDF of an asymmetric Laplace distribution. Plot (b) corresponds to probit regression with a non-linear systematic component. In the bottom row, (e) corresponds to a linear systematic component with an asymmetric inverse link function given by the CDF of a Gaussian distribution with standard deviation equal to a linear function of the covariate (i.e., $g^{-1}\lp\eta(x) = \beta_0 + \beta_1 x | \gamma_0, \gamma_1\rp = \Phi\lp\frac{\beta_0 + \beta_1 x}{\gamma_0 + \gamma_1 x}\rp$). Plot (f) is another instance of a probit link paired with a non-linear systematic component.

The important conclusion conveyed by these pairs of equivalent models is that asymmetry in a probability curve can arise for a variety of reasons including asymmetry, heteroskedasticity, or non-linearity (or some combination thereof) in the auxiliary variables. Moreover, certain characteristics of the auxiliary variables such as asymmetry and heteroskedasticity cannot be identified from the data without making strong assumptions about their relationship with the covariates.

Stated in terms of the three components of a GLM, the conclusion is that the link function and systematic component do not operate independently on the probability curve. Indeed, provided sufficient flexibility is permitted for the systematic component, $\eta(\mathbf{x})$, the particular choice of the link function has no impact on the flexibility of the resulting probability map $p$. Thus, without scientific knowledge that provides the basis for modeling assumptions such as linearity or a plausible auxiliary variable distribution, one defensible modeling approach has been to pair a convenient probit link function and a non-parametric systematic component \citep[e.g.,][]{Choudhuri2007}.

The examples in Figure~\ref{fig:matching_auxiliary} illustrate that relaxing assumptions about the conditional distribution of the auxiliary variables can be equivalent to relaxing assumptions of linearity. As the data are unable to help the researcher choose between the models associated with the first and second columns, the only reason to prefer one equivalent model over another would be to accommodate existing scientific knowledge about the data generating process. The example of ecological presence/absence data mentioned in Section~\ref{sec:auxiliary_var} presents one such case where acknowledging that the binary data arise due to thresholded observations of a discrete count process motivates the choice of a cloglog link (Poisson auxiliary variable) over a logistic one.

% Summarizing sentence:

\subsection{Quantile regression for binary data}\label{sec:QR}

One proposed approach for developing a flexible model for binary data uses quantile regression at the level of the auxiliary variable in a hierarchical model specification. When used in a continuous response setting, quantile regression is a linear model that relaxes distributional assumptions about the way the response varies around the linear trend. It is assumed that each quantile of the response distribution varies linearly with the covariates, but the linear effect of each covariate is permitted to vary across quantiles. The approach represents an effective way to model data exhibiting heteroskedastic and/or non-Gaussian residuals \citep{Koenker1978, Koenker2005}. An example of a heteroskedastic conditional random variable is shown in the bottom left plot of Figure~\ref{fig:matching_auxiliary}. The gray lines in Figure~\ref{fig:matching_auxiliary} show quantile curves for a selection of quantile levels. The parameters of interest in binary quantile regression are the slopes of these gray lines.

A generalized form of quantile regression has been proposed as a model for binary data by assuming that the responses, $y_i$, are generated from the auxiliary variable model in~\eqref{eqn:auxiliary} and only weak assumptions are made about the probability density, $f(z_i| \boldsymbol\beta)$ \citep{Benoit2012}. Namely, it is assumed that the quantiles of the auxiliary variables, $z_i$, exhibit so-called global linearity such that $Q(z_i| \boldsymbol\beta, \tau) = \mathbf{x}_i'\boldsymbol\beta(\tau)$ for all $\tau \in (0, 1)$, where $Q(z_i | \tau)$ denotes the $\tau\text{th}$ quantile of $z_i$ (this is true, for example, in plots (a) and (e) of Figure~\ref{fig:matching_auxiliary}). Analogous extensions have also been proposed for count-valued responses \citep{Machado2005, Lee2010, Williams2019}. In traditional logistic and probit regression, these quantiles are implicitly defined by specifying either a logistic or Gaussian density, respectively, and both cases assume homoskedasticity for the auxiliary variables. Quantile regression for binary data represents an attempt to estimate $\boldsymbol\beta(\tau)$ for a specific set of $\tau$, rather than define quantile curves through the specification of the density, $f(z_i| \boldsymbol\beta)$. A package for the R statistical programming language called \verb=bayesQR= \citep{Benoit2017} was recently developed that aims to provide practitioners with a tool to fit quantile regression models to data, including binary-valued data.

Beginning with the seminal work of \cite{Manski1985}, it has been known that quantile regression coefficients for a fixed quantile, $\tau$, are identifiable only up to a multiplicative constant when the data are binary. As we show below, the meaning of model parameters in binary quantile regression can easily be misinterpreted, and may not provide useful scientific learning when interpreted correctly. The results in \cite{Manski1985} are presented using algebraic arguments, whereas we present a geometric perspective that permits intuitive visualizations for the simplified cases of one or two covariates.

\subsubsection{Parameter identifiability: Single predictor}\label{sec:qr_identifiability1}

\begin{figure}[ht]
\centering
\includegraphics[width = \textwidth]{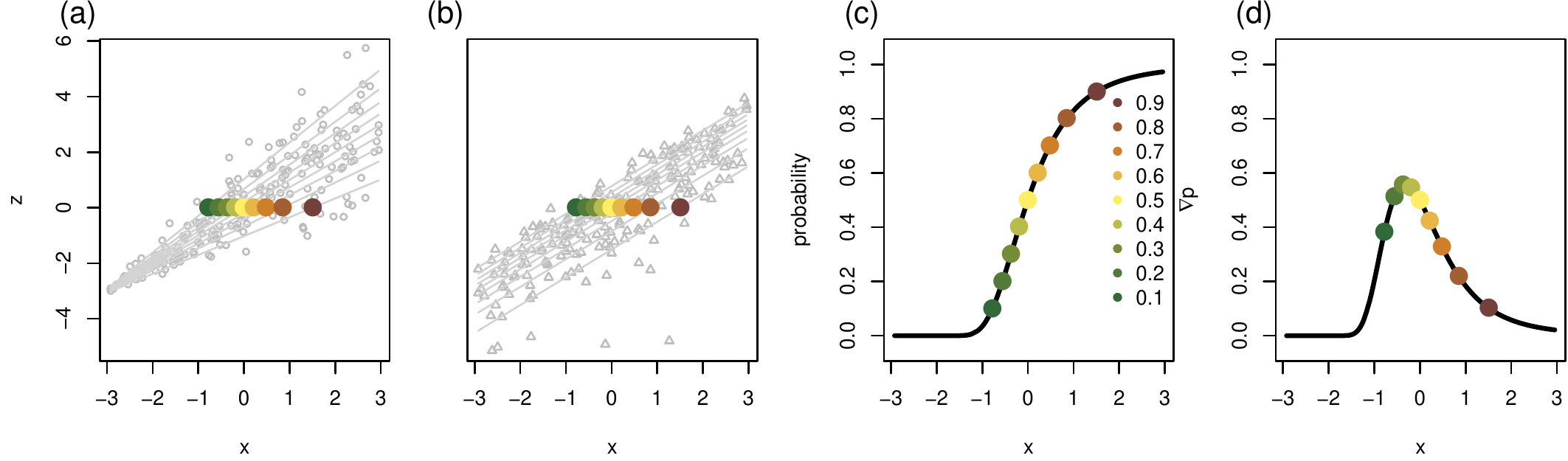}
\caption{Projection of quantiles onto intercepts of quantile functions for $z|x$. Plots (a) and (b) show realizations from two different auxiliary variable distributions that give rise to equivalent probability curves (c) and probability curve gradients (d).}
\label{fig:two_equiv_QR}
\end{figure}

Consider the probability curve for a hypothetical binary response that depends on an intercept and a single covariate, $x$, shown in plot (c) of Figure~\ref{fig:two_equiv_QR}, such that $\boldsymbol\beta(\tau) = (\beta_0(\tau), \beta_1(\tau))'$. A selection of nine quantiles are depicted as colored points along the curve. By definition, the value of $x$ at which the probability curve achieves a particular quantile, $\tau$, corresponds to the value of $x$ for which $Q(z | x, \beta_0, \beta_1, 1 - \tau) = 0$. That is, the proportion of the probability mass of the conditional random variable $z|x$ that occurs in the region $z > 0$ is $\tau$. Thus, a particular probability curve evaluated at a set of quantiles corresponds to a equal-sized set of points in the $xz$-plane at the locations where the quantile curves $Q(z|x, \beta_0, \beta_1, 1 - \tau)$ intersect the $x$-axis. Plot (a) and (b) in Figure~\ref{fig:two_equiv_QR} show two different possible distributions for $z | x$ that both have quantile curves intersecting the $x$-axis at the same locations, and both satisfy global linearity. Thus, they both give rise to exactly the same probability curve which implies the quantile regression parameters are not identifiable from any amount of binary data generated from that curve.

Importantly, the auxiliary variables in plot (a) exhibit heterogeneity in the covariate effects across quantiles, while the auxiliary variables in plot (b) do not. Hence, even under the global linearity assumption, it is not possible to identify the parameters $\boldsymbol\beta(\tau)$ from a binary response. It is not even possible to determine whether, under a global linearity assumption, the auxiliary variables exhibit heteroskedasticity of any kind. Geometrically, this is equivalent to stating that one cannot estimate the slope of the lines in the $xz$-plane; one can only estimate where they intersect the $x$-axis. For each fixed value of $\tau$, the non-identifiability of the slope is exactly the limitation pointed out in \cite{Manski1985}. The set of curves defined by $z = k\beta_0(\tau) + k\beta_1(\tau) x$ for $k \in \mathbb{R}$ coincides with the set of lines that pass through a common $x$-intercept, $\frac{-\beta_0(\tau)}{\beta_1(\tau)}$.

A natural question to ask is, why does the estimation procedure in \verb=bayesQR= appear to provide stable estimates of $\boldsymbol\beta(\tau)$? The answer lies in the nature of the inferential procedure implemented in the software package \verb=bayesQR=, which provides approximate inference about quantile-level effects by fitting a suite of sub-models to the data. In each sub-model, the auxiliary variables are assumed to arise from an asymmetric Laplace distribution with unit variance and known skewness parameter corresponding to one particular quantile of interest. Though all sub-models are almost surely misspecified, the procedure leads to valid posterior distributions for the covariate effects for continuous-response data in the sense that, for fixed $\boldsymbol\beta(\tau)$, posterior distributions constructed using this procedure converge asymptotically to point masses at the fixed levels \citep{Sriram2013, Yang2016}. However, posterior validity has not been established for the case of binary data.

Another way to characterize the identifiability problem with binary quantile regression is to state that the model is over-specified. As we noted in Section~\ref{sec:identifiability}, for a sufficiently flexible systematic component, $\eta(\mathbf{x})$, any probability curve can be defined by the map $g^{-1}(\eta(\mathbf{x}))$, regardless of the choice of link function $g$, or equivalently, the distribution of the auxiliary variables in a hierarchical model. Analogously, for a sufficiently flexible family of auxiliary variable distributions with parameters that may depend on the covariates, arbitrarily complex probability curves may be achieved for any fixed choice of the function $\eta(\mathbf{x})$. Quantile regression introduces the specific constraint of global linearity on the family of auxiliary variable distributions, but even with this enforced structure the assumptions of quantile regression are too weak to admit identifiable effects across quantiles.

By introducing constraints on the vector $\boldsymbol\beta(\tau)$, such as $||\boldsymbol\beta(\tau)||_2^2 = 1$ as suggested by \cite{Manski1985}, it is possible to identify certain functions of the quantile regression coefficients. The second plot in Figure~\ref{fig:two_equiv_QR} corresponds to a constraint that $\beta_1(\tau) = 1$ for all $\tau$. After a constraint is enforced to ensure identifiability, it is possible to use quantile regression methods to make inference about the probability curve \citep{Kordas2006}, although Bayesian methods that provide valid estimates of uncertainty have not yet been developed.

\subsubsection{Parameter identifiability: multiple predictors}\label{sec:qr_identifiability}
\begin{figure}[ht]
\centering
\includegraphics[width = 0.8\textwidth]{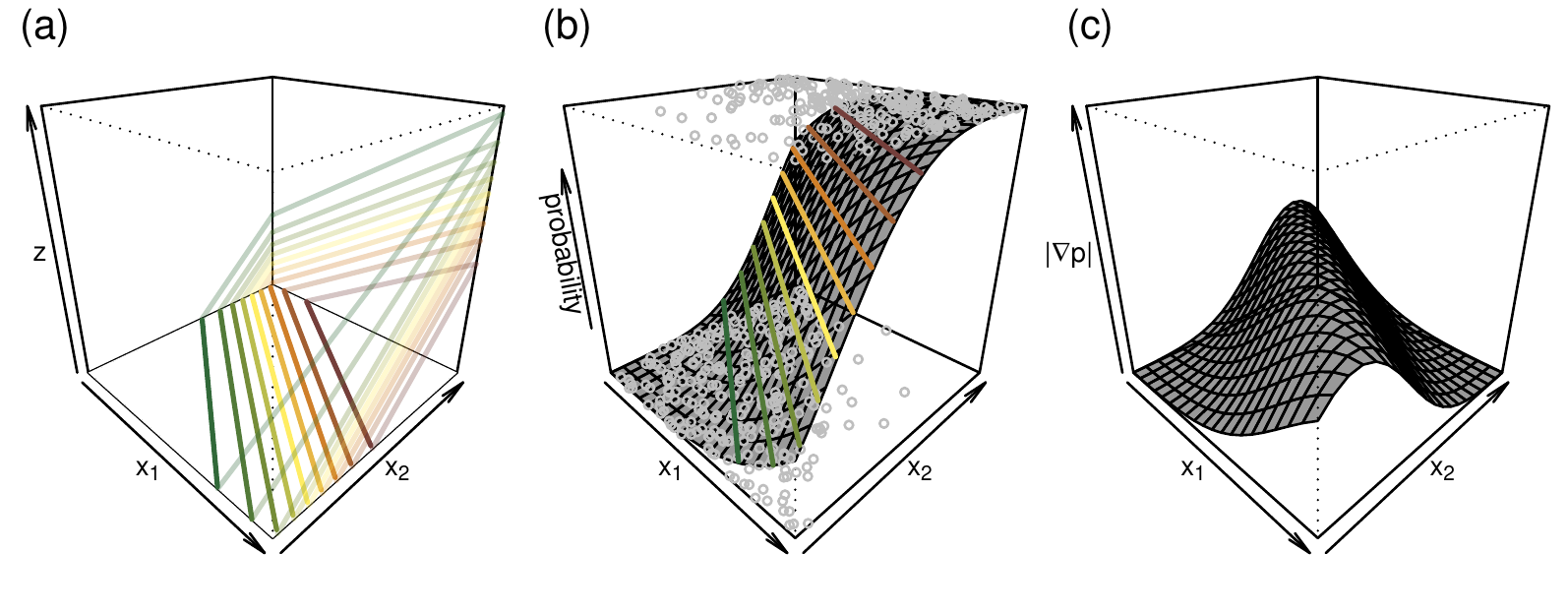}
\caption{Projection of quantiles onto intercepts of quantile functions for $z$. Plot (a) shows an example of auxiliary variables exhibiting global linearity that give rise to the probability surface in (b) and probability surface gradient, the magnitude of which is shown in (c).}
\label{fig:two_equiv_QR_MV}
\end{figure}

Typical regression analyses consider several covariates of interest. Figure~\ref{fig:two_equiv_QR_MV} shows how quantile surfaces for an auxiliary variable exhibiting global linearity with respect to two covariates (a) results in a particular probability surface (b) and corresponding gradient surface (c). The locations where the quantile curves intersect the two covariate axes are uniquely determined by the probability curve, as they were for the univariate case (Figure~\ref{fig:two_equiv_QR}). However, the slopes of the planes in the left plot are not fully identified without further constraints.

For the case of multiple predictors, a potentially useful parameter identifiable from the data is the relative effect of factor $j$ given by $\beta_j^*(\tau) = \beta_j(\tau) / |\boldsymbol\beta_{-0}(\tau)|$, where $|\boldsymbol\beta_{-0}(\tau)|$ is a suitable vector norm of all effects besides the intercept. This quantity can be interpreted as the contribution of the $j\text{th}$ covariate on the $\tau\text{th}$ quantile of the auxiliary variable relative to the total contributions of all covariates together. Thus, $\beta_j^*(\tau_1) < \beta_j^*(\tau_2)$, means that the relative contribution of the $j\text{th}$ factor is greater for quantile $\tau_2$ than for quantile $\tau_1$.

However, inequality between normalized effects tells us nothing about the change in the magnitude of an effect across different quantiles; it may be that $\beta_j(\tau_1) > \beta_j(\tau_2)$. Commonly produced ``forest plots'' that show how the value of $\beta_j(\tau)$ varies with $\tau$ are nonsensical in the context of a binary response because they present non-identifiable quantities. One could, in principle, examine $\beta_j^*(\tau)$ for a range of quantiles, $\tau$. However, the scientific learning afforded by measuring the relative contribution of a particular linear effect across quantiles is unclear. Lacking a coherent mechanistic or interpretive motivation, quantile regression for binary data reduces to another method like non-parametric probit regression that permits a high degree of flexibility in the shape of the probability curve. Other methods exist for estimating probability curves that offer equivalent levels of flexibility, yet do not rely on approximate inferential procedures \citep[e.g.,][]{Wood2017}. The necessity of quantile regression methods for binary data therefore remains to be established.

% Summarizing sentence:

\section{Modifying the random component}\label{sec:mod_random}

We have thus far introduced two ways to generalize GLMs for binary data to allow for different forms of flexibility in the probability surface. First, additional flexibility may be incorporated by relaxing the assumption of linearity in the systematic component of a traditional GLM and introducing higher order polynomial terms, or even taking a non-parametric perspective. Second, additional flexibility may be introduced by relaxing the structure of the link function, or equivalently, the distribution family of the conditional auxiliary variable in a hierarchical representation (Section~\ref{sec:mod_link}). A third way to introduce flexibility in the probability surface is to relax assumptions about the third component of a traditional GLM: the random component.

For binary data, the marginal distribution for the observations, $y_i$, must be Bernoulli. However, the tacit assumption made in traditional GLMs that the distributions of $y_i$ are conditionally independent can be relaxed through the introduction of random effects in the systematic component. For example, a simple extension of the probit regression model might allow for a random intercept in the systematic component, indexed by some underlying group structure, written mathematically as

\begin{align}
\begin{split}
  y_i | \boldsymbol\beta, \zeta_i &\sim \text{Bern}\lp p_i(\mathbf{x}_i, \boldsymbol\beta, \zeta_i) \rp, \\
  \Phi^{-1}(p_i) &= \mathbf{x}'_i \boldsymbol\beta + \zeta_i, \\
  \zeta_i &\sim \N \lp m(i), \sigma^2_\zeta \rp
\end{split}
\end{align}

where $m(i)$ is the intercept associated with the group to which datum $i$ belongs, and the intercept in the systematic component, $\beta_0$, is fixed at 0 to ensure parameter identifiability. Often there is no interest in estimating $\zeta_i$. Including the random intercept is done to ensure that any group-level effects that may be confounded with the covariates of interest are accounted for so that valid uncertainty estimates are produced for $\boldsymbol\beta$ (see \cite{Larsen2000} for a discussion of parameter interpretation in the analogous logistic model with random effects). A hierarchical formulation of this so-called generalized linear mixed model (GLMM) is given by

\begin{align}
\begin{split}
  y_i | z_i &= \mathbbm{1}_{z_i > 0}, \\
  z_i| \boldsymbol\beta, \zeta_i &\sim \N \lp \mathbf{x}'_i \boldsymbol\beta + \zeta_i, 1\rp, \\
  \zeta_i &\sim \N \lp m(i), \sigma^2_\zeta \rp.
\end{split}
\end{align}

Conditioned on the group random effect, $\boldsymbol\zeta = \lp \zeta_1, \dots, \zeta_n \rp'$, the data are independent as before, but integrating out the random effects induces positive dependence between random variables $y_i$ and $y_j | \mathbf{x}_j$ when $i$ and $j$ belong to the same group (i.e., $m(i) = m(j)$). Thus, the random component has been modified in its joint structure, although the marginal structure remains Bernoulli.

Random effects associated with other indices such as spatio-temporal information induce dependence for proximal responses. These types of models are useful when researchers suspect important explanatory variables correlated with the index are not included in $\mathbf{x}_i$. In Section~\ref{sec:application}, we demonstrate how a spatial random effect may be used to account for residual spatial structure in the occupancy pattern of European red squirrels (\textit{Sciurus vulgaris}) in Switzerland.

Issues with parameter identifiability can arise through the introduction of random effects in much the same way they do in traditional linear mixed models. Including $\zeta_i$ in the model increases the number of parameters to be estimated by $n$. In practice, this vector is typically assumed to exhibit strong positive dependence among variables so that the effective degrees of freedom will increase by an amount much less than $n$ and it will be possible to estimate the random effects from the data. In the example involving random intercepts, identifiability is achievable if the number of groups is much less than the number of observations. For spatial random effects, an analogous requirement is that the range of spatial dependence not be too short relative to the distances between spatial locations.

\section{Application to occupancy status of red squirrels}\label{sec:application}

We demonstrate how the concepts presented above on binary regression modeling may be used in practice with an application to the distribution of the European red squirrel in Switzerland. The data in this example are binary responses indicating whether any red squirrels were observed during visit $j$ to site $i$, and our goal was to develop a model for the occurrence of this species that allows us to infer the relationship between occurrence and relevant landscape covariates, as well as predict the occupancy status of new sites. The data were collected as part of the Swiss breeding bird survey (Monitoring H\"{a}ufige Brutv\"{o}gel, MHB; \citealt{Schmid2004}) carried out by the Swiss Ornithological Institute and previously analyzed using binary regression in \cite{Kery2015}. We developed two new models for occupancy that account for imperfect detection of the species and are parameterized by interpretable quantities related to species abundance. We compared the predictive performance of our proposed models with a baseline model that assumes constant occurrence and detection rates across sites and visits, in addition to a re-analysis based on the approach taken in \cite{Kery2015}.

\begin{figure}
\centering
\includegraphics[width = \textwidth]{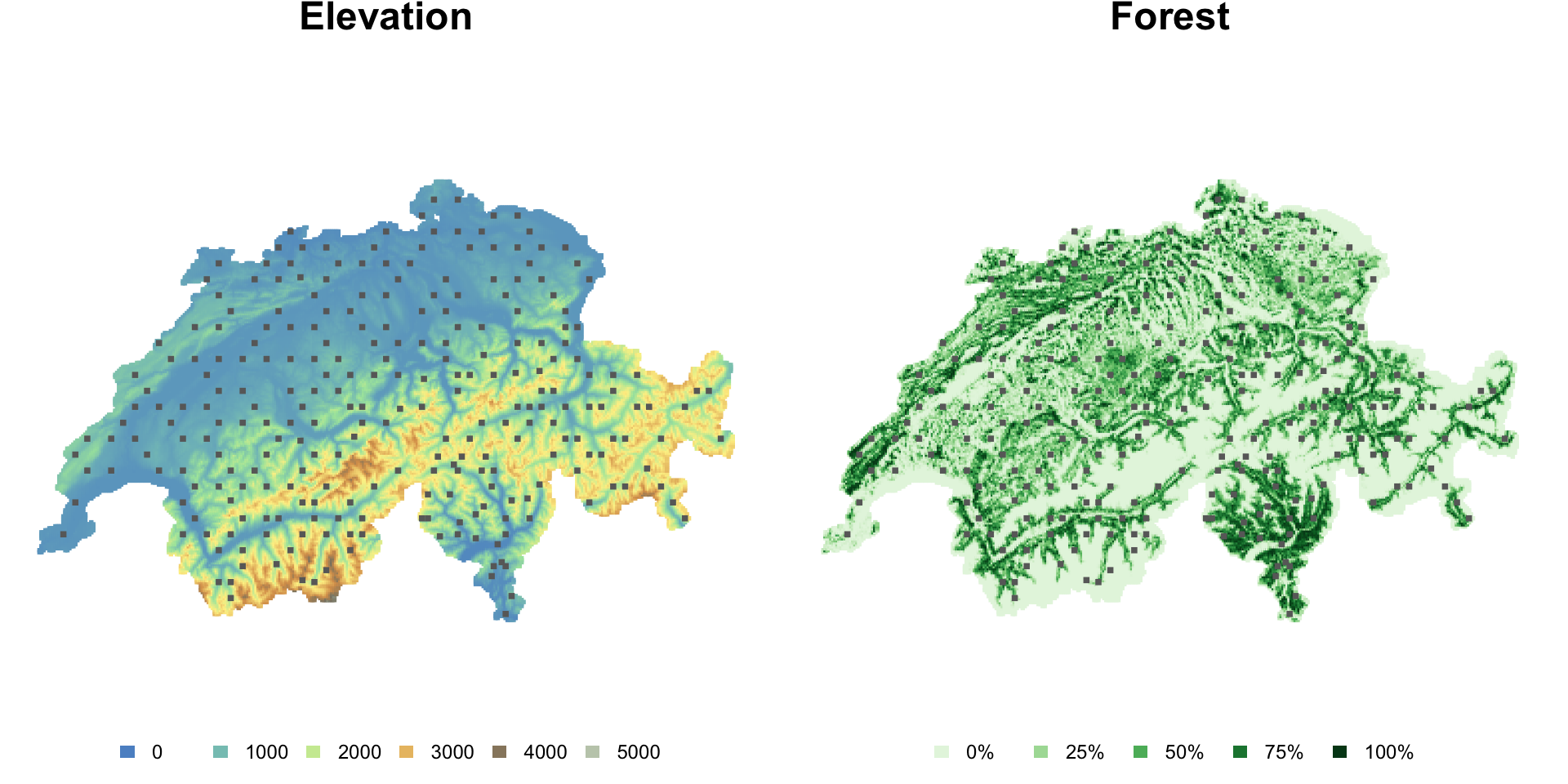}
\caption{Landscape covariates hypothesized to covary with occupancy.}
\label{fig:covariates}
\end{figure}

The 2007 data set is comprised of detections/nondetections (1/0) of the red squirrel in 265 1 km$^2$ survey quadrats in Switzerland. Each site was visited on as many as 3 separate occasions. Also recorded are several covariates of interest. Landscape-level covariates hypothesized to covary with occurrence probability ($\mathcal{X}$) are elevation and percent forest cover (see Figure~\ref{fig:covariates}). Covariates measured on each visit and hypothesized to covary with the detectability of red squirrels ($\mathcal{W}$) are the date and duration of the observation procedure, which ranged from April 13 to July 27 and 1.5 to 9.5 hours, respectively.

Setting aside the issue of detectability, in the presence/absence setting, a logistically- or normally-distributed auxiliary variable may not correspond to any observable natural phenomenon; however, as suggested in Section~\ref{sec:auxiliary_var}, a non-negative, integer-valued auxiliary variable could be interpreted as the true number of individuals present at a given site \citep[first proposed in][]{Royle2003}. If we let $z_i$ denote such a random variable for site $i$, then the probability of observing an individual under a scenario of perfect detection would be $\text{Pr}(z_i > 0)$. In two of the models we consider in this section, $z_i$ is modeled as a Poisson-distributed random variable with $\E{z_i | \boldsymbol\beta} = \lambda(\mathbf{x}_i) = e^{\mathbf{x}'_i \boldsymbol\beta}$.

To address the reality of imperfect detection, the probability of observing the study species at site $i$ on occasion $j$ is typically decomposed into the probability that the site is truly occupied, often denoted $\psi_i$, and the conditional probability that a species is detected, given the site is occupied, which we denote by $r_{ij}$. Allowing for imperfect detection modifies the model for the data such that $Pr(y_{ij} = 1 | \psi_i, r_{ij}) = \psi_i r_{ij}$. A hierarchical formulation for such an imperfect detection is given by

\begin{align}
\begin{split}
  y_{ij}|z_i, u_{ij} &= \mathbbm{1}_{z_i > 0}\mathbbm{1}_{u_{ij} > 0}, \\
  z_i| \boldsymbol\beta &\sim f_Z(\mathbf{x}_i, \boldsymbol\beta), \\
  u_{ij} | \boldsymbol\alpha &\sim
    f_U(\mathbf{w}_{ij}, \boldsymbol\alpha).
\end{split}
\end{align}

The occupancy probability $\psi_i$ is implicitly defined as the probability that $z_i$ is positive, and the conditional probability of detection, $r_{ij}$, is the probability that $u_{ij}$ is positive. A large body of literature has focused specifically on the development of statistical methods for occupancy data \citep[e.g.,][]{Mackenzie2002, MacKenzie2018a}, with emphasis on capturing temporal dynamics \citep[e.g.][]{Buckland1993, MacKenzie2003, Royle2007} and residual spatial dependence \citep[e.g.,][]{Buckland1993, Hooten2003, Johnson2013}.

There is clear motivation for specifying a discrete, non-negative distribution for $z_i$ such as the Poisson distribution. However, a natural motivation for the specification of $f_U$ is more illusive \citep[although see][for an example where detection probability was modeled as a function of abundance]{Royle2003}. It is reasonable to expect that the detectability of a species might depend on some relevant site- and visit-specific covariates. In the absence of a naturally arising distribution for $f_U$, we specified a standard normal distribution with mean $\mathbf{w}_{ij}'\boldsymbol\alpha$, where $\mathbf{w}_{ij}$ is a vector of covariates related to detectability at site $i$ on visit $j$ \citep[as in][]{Dorazio2012}.

We considered a total of four model formulations outlined below. Further details including prior specifications, and algorithms used to sample from the joint posterior distributions are included in the supplementary materials. The most basic, ``naive'' model we considered assumes that the probabilities of occupancy and detection (given a site is occupied) are constant across both sites and visits:

\begin{align}
\begin{split}
  y_{ij} &\sim \text{Bern}\lp r, \psi \rp, \\
  r &\sim \text{Beta} \lp a_r, b_r \rp, \\
  \psi &\sim \text{Beta} \lp a_\psi, b_\psi \rp.
\end{split}
\end{align}

A second model extends traditional logistic regression for the case of occupancy data with imperfect detection. This model is presented in \cite{Kery2015}, and referred to here as the K\'{e}ry-Royle model:

\begin{align}
\begin{split}
  y_{ij} | z_i, u_{ij} &= \mathbbm{1}_{z_i > 0}\mathbbm{1}_{u_{ij} > 0}, \\
  z_i | \boldsymbol\beta &\sim \text{Logistic} \lp \mathbf{x}_i'\boldsymbol\beta, 1 \rp, \\
  u_{ij} | \boldsymbol\alpha &\sim
    \text{Logistic}\lp \mathbf{w}_{ij}'\boldsymbol\alpha, 1 \rp.
\end{split}
\end{align}

A third model assumes Poisson-distributed auxiliary variables related to occupancy, and normally distributed auxiliary variables for detection:

\begin{align}
\begin{split}
  y_{ij} | z_i, u_{ij} &= \mathbbm{1}_{z_i > 0}\mathbbm{1}_{u_{ij} > 0}, \\
  z_i| \boldsymbol\beta &\sim
    \text{Pois}\lp \lambda(\mathbf{x}_i) = e^{\mathbf{x}_i'\boldsymbol\beta}\rp, \\
  u_{ij} | \boldsymbol\alpha &\sim
    \N \lp \mathbf{w}_{ij}'\boldsymbol\alpha, 1 \rp.
\end{split}
\end{align}

Finally, a fourth model extends the Poisson model by introducing a spatial random effect:

\begin{align}
\begin{split}
  y_{ij} | z_i, u_{ij} &= \mathbbm{1}_{z_i > 0}\mathbbm{1}_{u_{ij} > 0}, \\
  z_i | \boldsymbol\beta, \zeta_i &\sim
    \text{Pois}\lp \lambda(\mathbf{x}_i) = e^{\mathbf{x}_i'\boldsymbol\beta + \zeta_i}\rp, \\
  u_{ij} | \boldsymbol\alpha &\sim
    \N \lp \mathbf{w}_{ij}'\boldsymbol\alpha, 1 \rp
\end{split}
\end{align}

The spatial random effect, $\zeta_{i}$, is modeled as a Gaussian random vector with conditionally Markovian dependence structure. There is a considerable literature concerned with the development of spatial multivariate Gaussian distributions \citep[e.g.,][]{Cressie1991, Rue2005a, Gelfand2010}. For the purposes of illustration, we set aside investigation of competing spatial models for the random effect and considered only Gaussian Markov random fields with a known neighborhood structure \citep{Rue2005a}.

We specified a first-order conditional autoregressive \citep[CAR; e.g.,][]{VerHoef2018} distribution for $\boldsymbol\zeta$, such that the precision matrix is defined as $\mathbf{Q} = \tau^2 \lp \mathbf{D} - \rho \mathbf{A} \rp$, where $\mathbf{A}$ is a binary adjacency matrix representing a known neighborhood structure, and $\mathbf{D}$ is diagonal with entries equal to the sum of the neighbors for each site (i.e., the row sums of $\mathbf{A}$). We defined $\mathbf{A}$ as the adjacency matrix of sites in the unique Delauney triangulation of the observation locations. The single parameter, $\rho$, controls the strength of the conditional dependence among neighbors so that as $\rho \rightarrow 1$, $\boldsymbol\zeta$ becomes constant-valued across sites, and as $\rho \rightarrow 0$, $\boldsymbol\zeta$ becomes a collection of independent, identically distributed Gaussian random variables. Thus, requiring that $\rho$ is sufficiently far from 0 is one way to ensure identifiability of the spatial random process.

The spatial random effect could, in principle, be introduced at one of several locations in the hierarchical structure, but the most natural place is at the same level as the occupancy covariates (\citealt{Johnson2013}, and see \citealt{Schmidt2015} for an approach that allows for non-stationarity). Introducing the random effect in this way is convenient because $\zeta_i$ has support given by the entire real line, and $e^{\zeta_i}$ may be interpreted as the extra multiplicative effect on the expected abundance at site $i$ due to unobserved covariates.

For each of the K\`{e}ry-Royle (KR), Poisson (P), and spatial-Poisson (SP) models, we considered two possible collections of covariates for both $\mathcal{X}$ and $\mathcal{W}$. The first collection includes the first-order effects of elevation and percent forest cover on the auxiliary variable related to occupancy ($z_i$), and the first-order effects of date and duration on the auxiliary variable related to detection ($u_{ij}$). The second collection includes all the first-order effects related to occupancy, as well as the quadratic effects of elevation and percent forest cover (elevation$^2$ and forest$^2$), the interaction between elevation and percent forest cover (elevation$\times$forest), the interaction between the quadratic effect of elevation and linear effect of percent forest cover (elevation$^2 \times$forest), and the interaction between the linear effect of elevation and quadratic effect of percent forest cover (elevation$\times$forest$^2$). The second collection also includes the first-order effects of date and duration, as well as the quadratic effect of duration (duration$^2$) related to detection. We refer to the first collection as linear effects (L) and the second collection as the quadratic effects (Q) (see Table~\ref{tab:covariate_collections}). The second collection was selected based on the analysis of \cite{Kery2015}. Matching the proposed model structures with the two collections of covariates results in seven predictive models for the occurrence of red squirrels: the naive model (N), two K\'{e}ry-Royle models (KRL and KRQ), two Poisson models (PL and PQ), and two spatial Poisson models (SPL and SPQ).

\begin{table}
\caption{Collections of covariates}
\label{tab:covariate_collections}
\centering
  \begin{tabular}{rcc|rcc}
    \multicolumn{3}{c|}{occupancy ($\mathcal{X}$)} &
      \multicolumn{3}{c}{detection ($\mathcal{W}$)} \\
    \hline
    covariate & L & Q & covariate & L & Q\\
    \hline
    intercept & \checkmark & \checkmark &
      intercept & \checkmark & \checkmark \\
    elevation & \checkmark & \checkmark &
      date & \checkmark & \checkmark \\
    forest & \checkmark & \checkmark &
      duration & \checkmark & \checkmark \\
    elevation$^2$ & & \checkmark &
      duration$^2$ & & \checkmark \\
    forest$^2$ & & \checkmark &
      & & \\
    elevation$^2 \times$ forest & & \checkmark &
      & & \\
    elevation $\times$ forest$^2$ & & \checkmark &
      & &
  \end{tabular}
\end{table}

Before discussing the results of each proposed model/covariates combination, we briefly summarize what practical and scientific considerations a researcher might use to select from among these possibilities, setting aside the naive model as an overly simplistic approach used only as a baseline against which we compare the performance of more plausible models. The KR approach represents the most traditional approach from the literature for analyzing data of this type. Points in favor are the existence of vetted model fitting software and the ability to include non-linear effects of covariates on both occupancy and detection probabilities, yielding flexible probability curves for the data. The primary limitations of the KR models are that neither the auxiliary variables, nor the effects parameters, $\boldsymbol\beta$, are directly interpretable. If the primary scientific goals include only prediction and estimation of the probability curves for occupancy and detection, then the KR approach represents a valuable and appropriate statistical tool.

The Poisson-based approaches provide the same level of flexibility for the occupancy and detection probability curves, with the additional benefit of interpretable auxiliary variables. Using one of these approaches, one could, for example, obtain an estimate with uncertainty of the abundance of red squirrels at any of the visited sites, or predict abundance at a new site. Though we note that such abundance estimates can depend strongly on the particular non-negative discrete-valued distribution specified for $z_i| \boldsymbol\beta$, \citep[\textit{sensu}][]{Barker2017}, they can nevertheless provide preliminary estimates useful, for example, in developing future study designs. The spatial Poisson approach provides even greater flexibility for the probability curves of occupancy and detection and allows for the possibility that important landscape-level covariates may be missing from the analysis. The primary drawback of the Poisson approaches are that they require a greater investment in the development of custom model fitting software, although user-friendly systems for fitting custom Bayesian models to data such as NIMBLE \citep{DeValpine2017} minimize the impact of this challenge. In addition, introducing a spatial random effect necessitates a more sophisticated investigation on the part of the analyst into the existence of weakly-identified model parameters and overall goodness of model fit.

\subsection{Results}

\begin{figure}
\centering
\includegraphics[width = \textwidth]{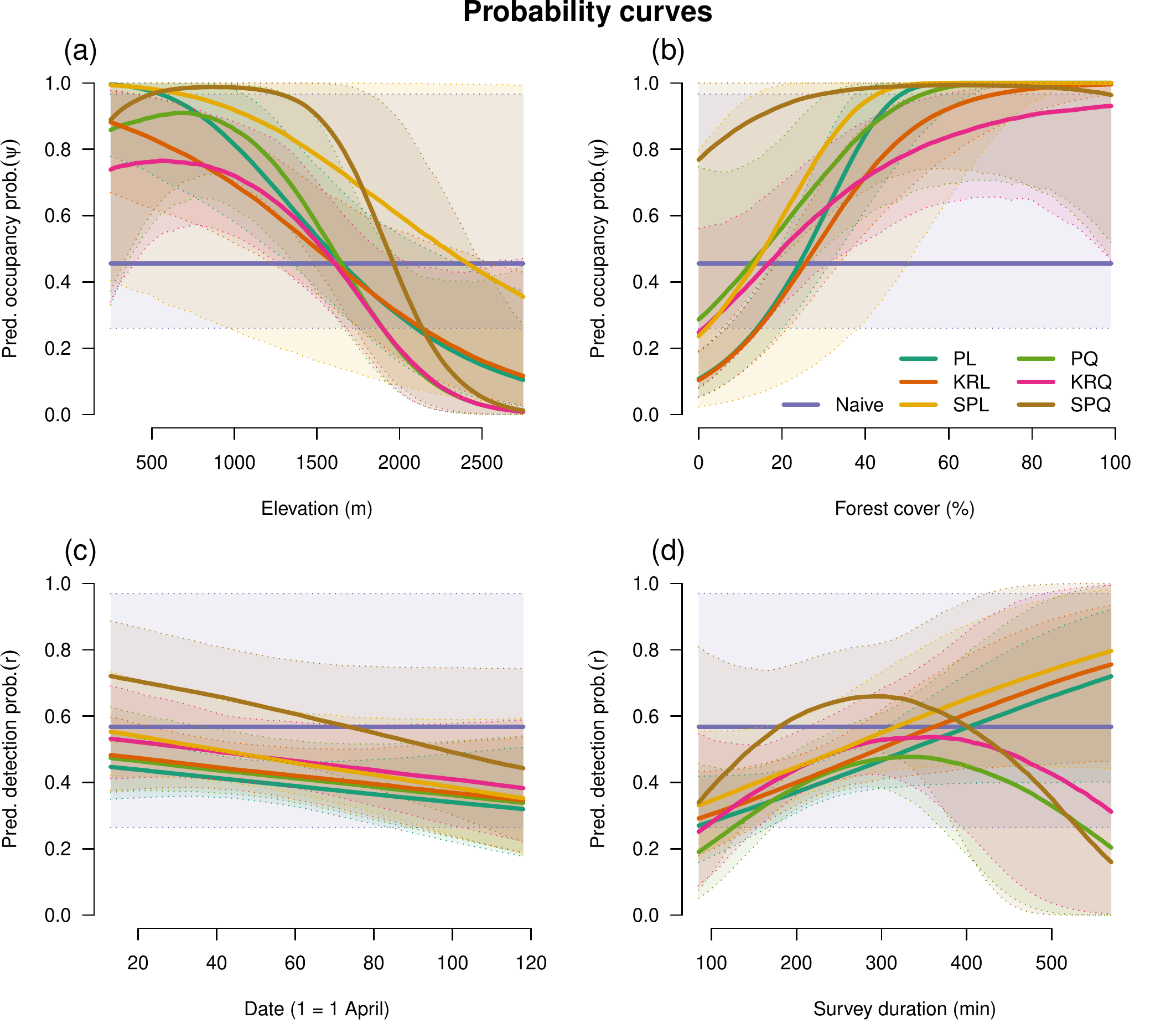}
\caption{Marginal probability curves for the probability of occupancy as a function of elevation (a) and percent forest cover (b), and the conditional probability of detection given occupancy as a function of survey date (c) and survey duration (d). Colored lines represent point-wise posterior medians from each model, and dotted lines give point-wise equal-tailed 95\% credible intervals.}
\label{fig:effects}
\end{figure}

Figure~\ref{fig:effects} shows the marginal probability curves for occupancy as functions of elevation (a) and percent forest cover (b) for all seven model-covariate combinations, and the marginal probability curves for detection as a functions of visit date (c) and duration (d). The overall shapes of the curves are consistent across the six non-naive models, with the possible exception of detectability as a function of survey duration for which models incorporating quadratic covariates show a decreasing effect of survey duration on detectability above 6 hours, albeit with a considerable increase in uncertainty compared to the linear covariates.

\begin{figure}
\centering
\includegraphics[width = \textwidth]{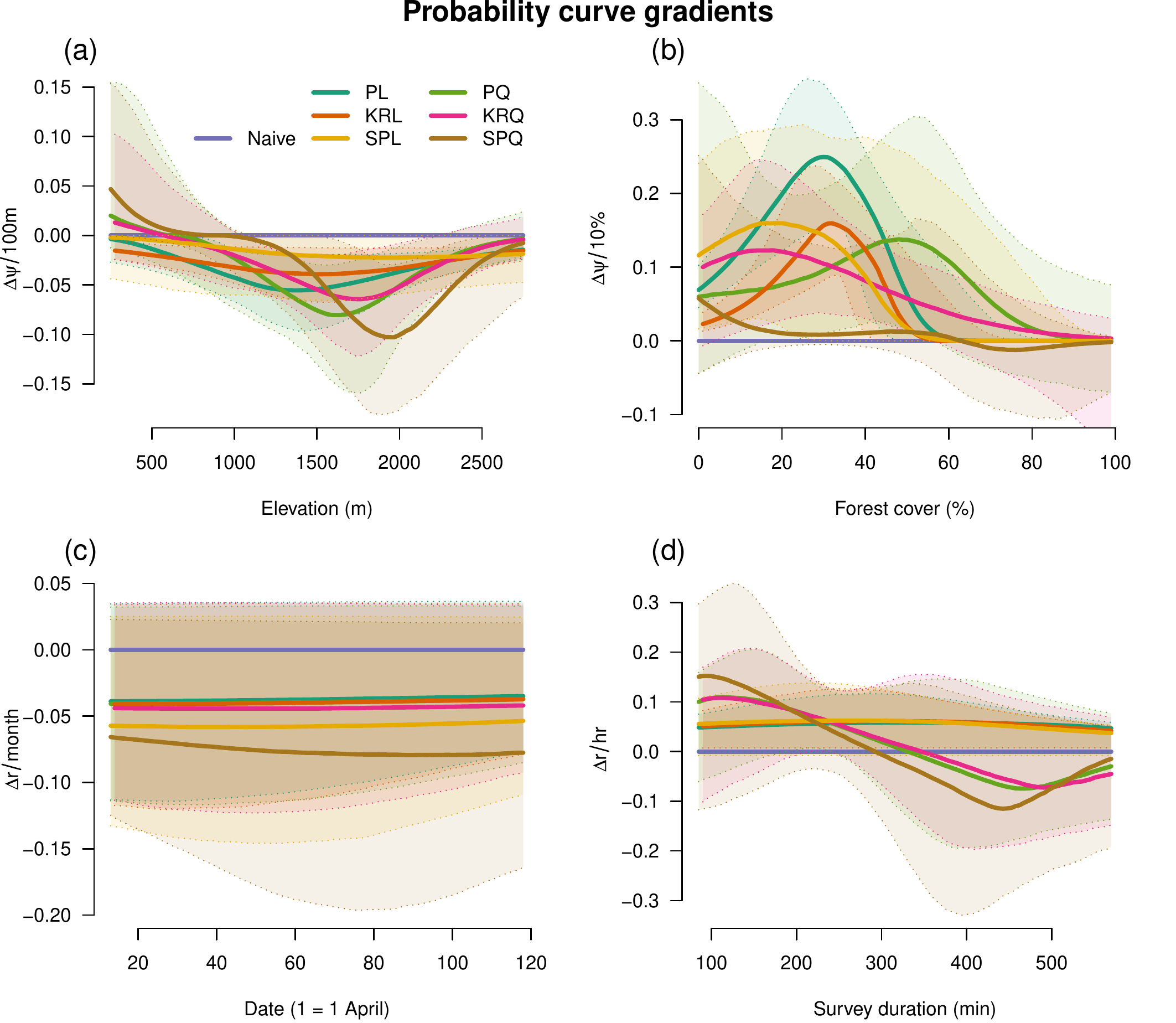}
\caption{Derivatives of marginal probability curves in Figure~\ref{fig:effects} arranged analogously. Colored lines represent point-wise posterior medians from each model, and dotted lines give point-wise equal-tailed 95\% credible intervals.}
\label{fig:grad_effects}
\end{figure}

Figure~\ref{fig:grad_effects} shows the marginal probability gradients, arranged analogously to Figure~\ref{fig:effects}. From plot (b) in Figure~\ref{fig:grad_effects}, we can see that a notable difference between the PQ (green) and KRQ (pink) models is in what region of percent forest cover the occupation probabilities are most sensitive to changes in forest cover, with the Poisson model suggesting peak sensitivity at higher levels of percent forest cover.

\begin{figure}
\centering
\includegraphics[width = \textwidth]{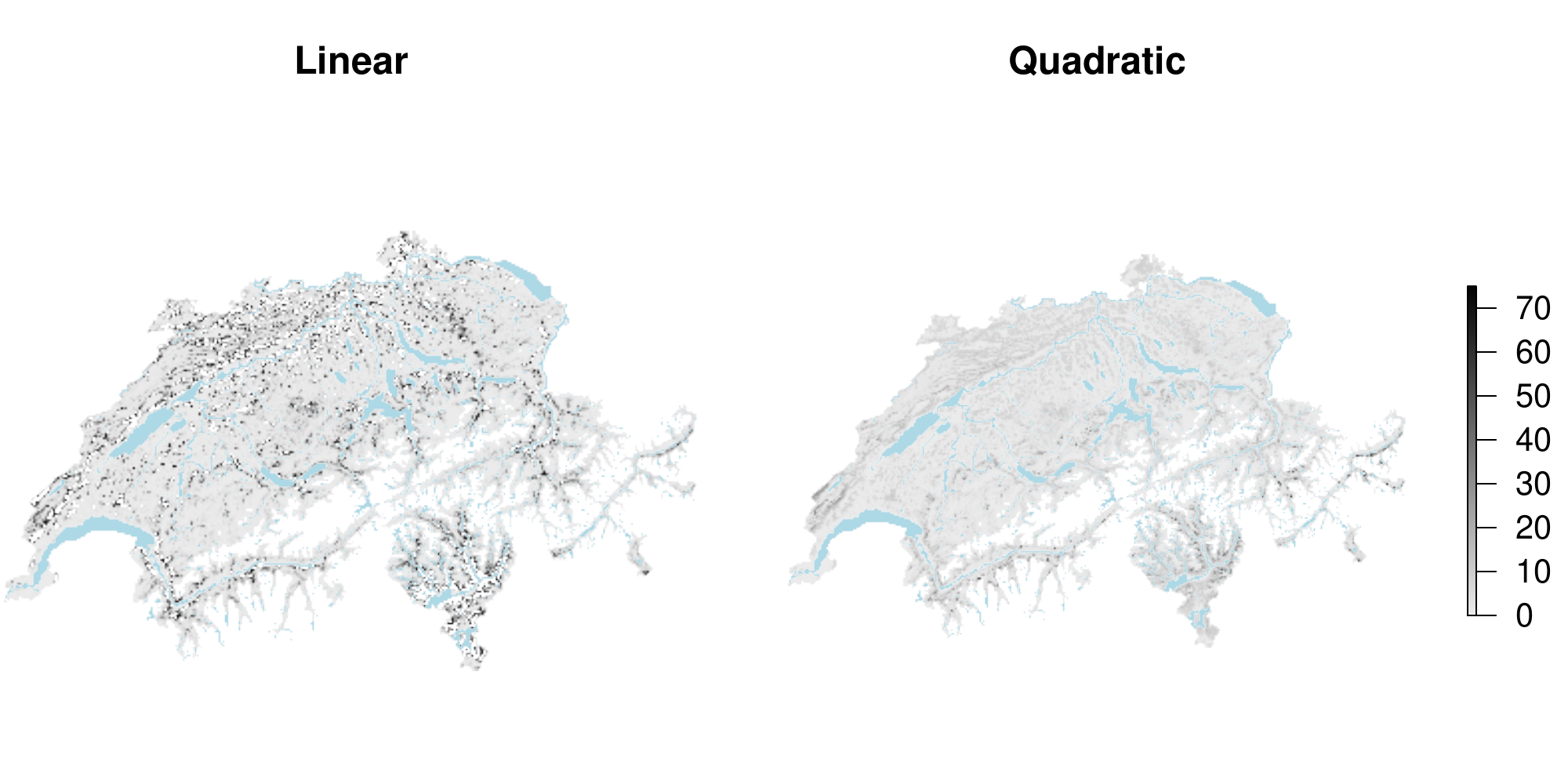}
\caption{Posterior median of abundance intensity, $\lambda$, for the PL and PQ models. Light blue represents major bodies of water.}
\label{fig:lambda}
\end{figure}

The Poisson-based approaches allow for abundance estimates to be made at both the surveyed sites and new locations. Figure~\ref{fig:lambda} shows the posterior median of the abundance intensity, $\lambda$, across Switzerland at a resolution of 1km$^2$. A notable difference between the linear and quadratic covariate combinations is the volatility of the abundance intensity estimates. The range of median abundance intensity across Switzerland is much larger for PL (95\% of locations have posterior median abundance intensity less than 74) than for PQ (95\% of locations less than 10), and the variability in abundance intensity between neighboring locations is also much larger for PL (median difference in posterior median abundance intensity of 4.4) than PQ (1.3). In summary, the quadratic covariates predict a smoother abundance surface than the linear one.

\begin{figure}
\centering
\includegraphics[width = \textwidth]{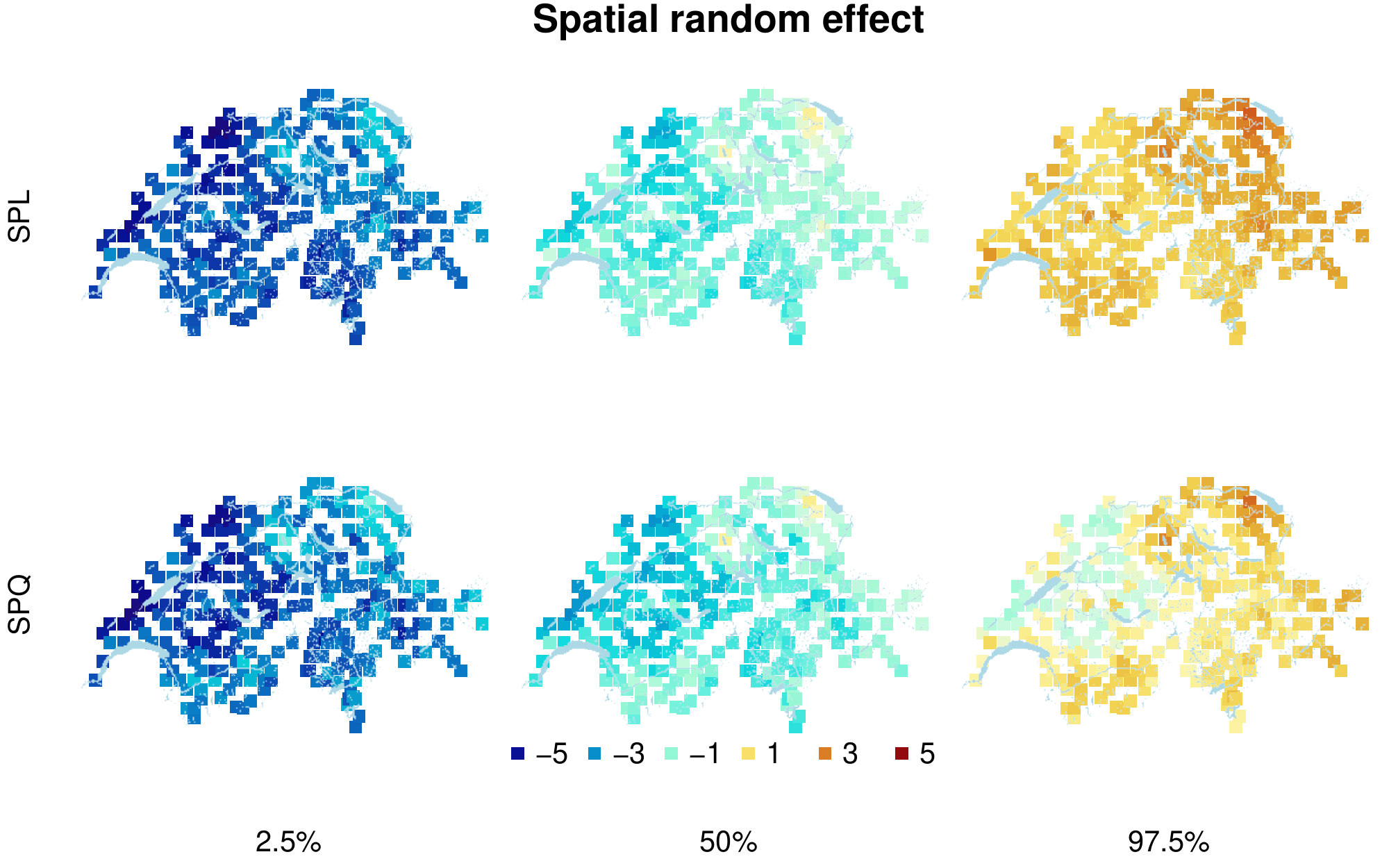}
\caption{Posterior summary of spatial random effect, $\boldsymbol\zeta$, for the SPL and SPQ models.}
\label{fig:eta}
\end{figure}

The spatial random effect in the SPL and SPQ models provides some insight into where the measured covariates alone may not be able to adequately explain variability in the data. Figure~\ref{fig:eta} shows three quantiles (0.025, 0.5, 0.975) summarizing the posterior distribution of the spatial random effect, $\zeta$, at each surveyed location. The random effect associated with the SPL model fit is, in general, less variable than the random effect associated with the SPQ model, as evidenced by the narrower range of values in the posterior median of $\zeta$ (e.g., -2.65 to 0.57 in the posterior median of $\zeta$ for SPL, compared to -3.06 to 0.63 for SPQ). Visually, the maps for SPL (top row Figure~\ref{fig:eta}) appear more constant than those for SPQ. The posterior median of $\zeta$ for SPQ (middle plot, bottom row Figure~\ref{fig:eta}) suggests there may be characteristics in the western part of Switzerland not captured by the covariates that reduce the probability of occupancy.

\begin{figure}
\includegraphics[width = 0.75\textwidth]{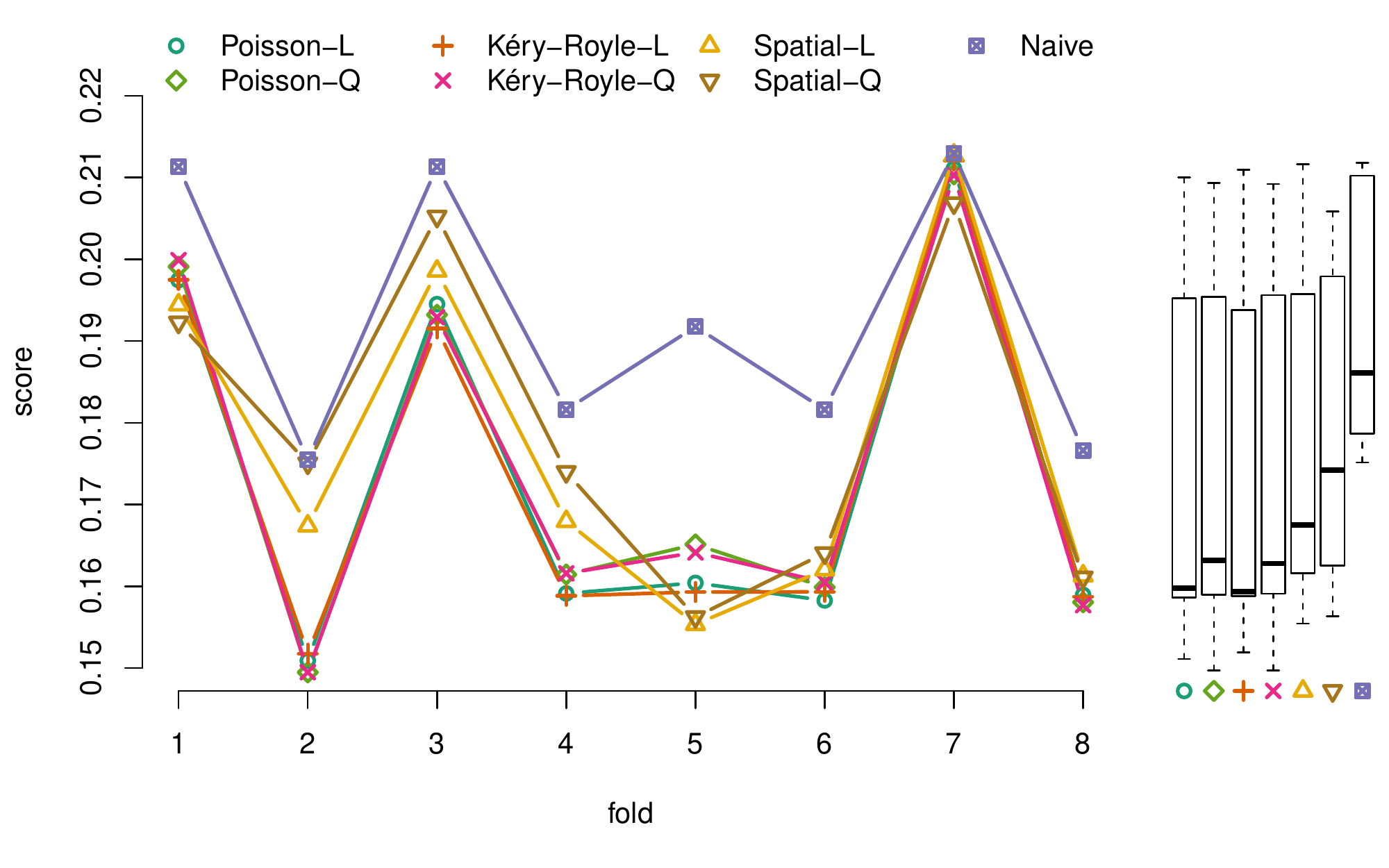}
\caption[]{K-fold cross validation scores indexed by holdout set (left), and boxplots aggregating across holdout sets (right).}
\label{fig:CV_scores}
\end{figure}

We also investigated the predictive performance of each of the seven models. We used a $K$-fold cross validation approach and scored each hold out set, $s_k$, using the proper score

\begin{align}
  S(\mathbf{y}_{s_k}, \hat{\mathbf{p}}_{s_k}) = |s_k|^{-1} \sum_{i \in s_k} (y_i - \hat{p}_i)^2,
\end{align}

where $|s_k|$ is the size of the $k\text{th}$ of $K$ holdout sets, and $\hat{p}_i$ is a predicted probability of observing a red squirrel at site $i$. To capture the uncertainty in the predicted probabilities, $p_i$, we computed $S(\mathbf{y}_{s_k}, \hat{\mathbf{p}}_{s_k})$ for every realization from the posterior distribution of $p_i$ for each model and used the sample average as an estimate of the expected value of the score with respect to the posterior distribution. Figure~\ref{fig:CV_scores} shows the expected score for each model across $K=8$ folds chosen completely at random. The only clearly inferior model is the naive one. All other models are arguably indecipherable from one another in terms of predictive performance. Thus, there does not appear to be sufficient information in the data to select from the plausible models presented. A researcher would therefore need to consider the benefits of extra interpretability afforded by the Poisson-based models, the availability of statistical software to perform model fitting, and whether the model might suffer from issues of identifiability or poor fit \citep[see also][for arguments in favor of selecting interpretable models even when predictive scores are worse]{VerHoef2015}.

\section{Discussion}

Flexible and interpretable regression models for binary data are in continual demand across a wide variety of scientific disciplines. After first briefly reviewing conventional approaches for binary regression in common use, we showed how new and existing binary regression models can be constructed using auxiliary variables. Hierarchical binary regression models specified using auxiliary variables have the benefit of facilitating the creation of models with interpretable parameterizations when the auxiliary variables correspond to a real but unobserved natural process that gives rise to the observed binary response. Because binary data can often be understood as summaries of natural processes through thresholding or censorship, interpretable auxiliary variables are often intuitive to define.

A necessary step in the development of any new statistical methodology, establishing parameter identifiability in a proposed binary regression model is of critical importance and not always straightforward. In the pursuit of flexible models for data that require minimal assumptions, it is possible to inadvertently introduce non-identifiability issues. Approximate model fitting procedures can sometimes mask issues where fully Bayesian inference might be more revealing. Unfortunately, exact inferential procedures are not always available.

Generalizations to traditional binary regression models can be usefully characterized by the ways in which modifications are made to the three components of generalized linear models (random component, systematic component, and link function). Through an application to the occurrence of red squirrels in Switzerland, we demonstrated how a researcher might construct several potential hierarchical models for binary data. The occupancy status of an areal unit can be thought of as a thresholded observation of abundance at 0, which motivated the specification of a Poisson-distributed auxiliary variable corresponding to site abundance. This auxiliary variable distribution implied a modification of the traditional symmetric link function to the complementary log-log link function. We also proposed a further generalization of this latent abundance model by including a spatial random effect, which implied a modification of the traditional assumption of conditional independence in the random component.

One avenue for future research that could yield more reliable quantile regression inference for binary data would be to extend one of the more holistic Bayesian approaches of \cite{Kottas2009, Taddy2010, Reich2011, Tokdar2012, Yang2017} to the case of binary data. Although many of the same identifiability issues that appear in existing quantile regression approaches for binary data would remain, these alternative methods do not rely on the approximate inferential procedure employed in the \verb=R= package \verb=bayesQR=. Fully Bayesian inference would aid in the detection of identifiability issues and obviate the need for theoretical guarantees surrounding approximation error.

\section{Acknowledgments}

Any use of trade, firm, or product names is for descriptive purposes only and does not imply endorsement by the U.S. Government.

\appendix

\clearpage
\bibliographystyle{biom}
\bibliography{HBR_review.bib}

\begin{thebibliography}{}

\bibitem[\protect\citeauthoryear{Agresti}{Agresti}{2002}]{Agresti2002}
Agresti, A. (2002).
\newblock {\em {Categorical Data Analysis}}.
\newblock John Wiley \& Sons, Hoboken, NJ, 2nd edition.

\bibitem[\protect\citeauthoryear{Albert and Chib}{Albert and
  Chib}{1993}]{Albert1993}
Albert, J.~H. and Chib, S. (1993).
\newblock {Bayesian analysis of binary and polychotomous response data}.
\newblock {\em Journal of the American Statistical Association} {\bf 88,}
  669--679.

\bibitem[\protect\citeauthoryear{Augustin, Mugglestone, and Buckland}{Augustin
  et~al.}{1996}]{Augustin1996}
Augustin, N.~H., Mugglestone, M.~A., and Buckland, S.~T. (1996).
\newblock {An autologistic model for the spatial distribution of wildlife}.
\newblock {\em Journal of Applied Ecology} {\bf 33,} 339--347.

\bibitem[\protect\citeauthoryear{Barker, Schofield, Link, and Sauer}{Barker
  et~al.}{2018}]{Barker2017}
Barker, R.~J., Schofield, M.~R., Link, W.~A., and Sauer, J.~R. (2018).
\newblock {On the reliability of N-mixture models for count data}.
\newblock {\em Biometrics} {\bf 74,} 369--377.

\bibitem[\protect\citeauthoryear{Baz{\'{a}}n, Bolfarine, and
  Branco}{Baz{\'{a}}n et~al.}{2010}]{Bazan2010}
Baz{\'{a}}n, J.~L., Bolfarine, H., and Branco, M.~D. (2010).
\newblock {A framework for skew-probit links in binary regression}.
\newblock {\em Communications in Statistics -- Theory and Methods} {\bf 39,}
  678--697.

\bibitem[\protect\citeauthoryear{Benoit and {Van den Poel}}{Benoit and {Van den
  Poel}}{2012}]{Benoit2012}
Benoit, D.~F. and {Van den Poel}, D. (2012).
\newblock {Binary quantile regression: a Bayesian approach based on the
  asymmetric Laplace distribution}.
\newblock {\em Journal of Applied Econometrics} {\bf 27,} 1174--1188.

\bibitem[\protect\citeauthoryear{Benoit and {Van den Poel}}{Benoit and {Van den
  Poel}}{2017}]{Benoit2017}
Benoit, D.~F. and {Van den Poel}, D. (2017).
\newblock {bayesQR: A Bayesian approach to quantile regression}.
\newblock {\em Journal of Statistical Software} {\bf 76,} 1--32.

\bibitem[\protect\citeauthoryear{Buckland and Elston}{Buckland and
  Elston}{1993}]{Buckland1993}
Buckland, S.~T. and Elston, D. (1993).
\newblock {Empirical models for the spatial distribution of wildlife}.
\newblock {\em Journal of Applied Ecology} {\bf 30,} 478--495.

\bibitem[\protect\citeauthoryear{Chen, Dey, and Shao}{Chen
  et~al.}{1999}]{Chen1999}
Chen, M.~H., Dey, D.~K., and Shao, Q.~M. (1999).
\newblock {A new skewed link model for dichotomous quantal response data}.
\newblock {\em Journal of the American Statistical Association} {\bf 94,}
  1172--1186.

\bibitem[\protect\citeauthoryear{Choudhuri, Ghosal, and Roy}{Choudhuri
  et~al.}{2007}]{Choudhuri2007}
Choudhuri, N., Ghosal, S., and Roy, A. (2007).
\newblock {Nonparametric binary regression using a Gaussian process prior}.
\newblock {\em Statistical Methodology} {\bf 4,} 227--243.

\bibitem[\protect\citeauthoryear{Conn, Johnson, Williams, Melin, and
  Hooten}{Conn et~al.}{2018}]{Conn2018}
Conn, P.~B., Johnson, D.~S., Williams, P.~J., Melin, S.~R., and Hooten, M.~B.
  (2018).
\newblock {A guide to Bayesian model checking for ecologists}.
\newblock {\em Ecological Monographs} {\bf 88,} 526--542.

\bibitem[\protect\citeauthoryear{Cressie}{Cressie}{1991}]{Cressie1991}
Cressie, N. (1991).
\newblock {\em {Statistics for Spatial Data}}.
\newblock John Wiley \& Sons, New York, New York, USA, first edition.

\bibitem[\protect\citeauthoryear{de~Valpine, Turek, Paciorek, Lang, and
  Bodik}{de~Valpine et~al.}{2017}]{DeValpine2017}
de~Valpine, P., Turek, D., Paciorek, C.~J., Lang, D.~T., and Bodik, R. (2017).
\newblock {Programming with models: Writing statistical algorithms for general
  model structures with NIMBLE}.
\newblock {\em Journal of Computational and Graphical Statistics} {\bf 26,}
  403--417.

\bibitem[\protect\citeauthoryear{Diggle, Tawn, and Moyeed}{Diggle
  et~al.}{1998}]{Diggle1998}
Diggle, P.~J., Tawn, A.~J., and Moyeed, R.~A. (1998).
\newblock {Model-based geostatistics}.
\newblock {\em Journal of the Royal Statistical Society: Series C (Applied
  Statistics)} {\bf 47,} 299--350.

\bibitem[\protect\citeauthoryear{Dorazio and Rodr{\'{i}}guez}{Dorazio and
  Rodr{\'{i}}guez}{2012}]{Dorazio2012}
Dorazio, R.~M. and Rodr{\'{i}}guez, D.~T. (2012).
\newblock {A Gibbs sampler for Bayesian analysis of site-occupancy data}.
\newblock {\em Methods in Ecology and Evolution} {\bf 3,} 1093--1098.

\bibitem[\protect\citeauthoryear{Fahrmeir and Raach}{Fahrmeir and
  Raach}{2007}]{Fahrmeir2007}
Fahrmeir, L. and Raach, A. (2007).
\newblock {A Bayesian semiparametric latent variable model for mixed
  responses}.
\newblock {\em Psychometrika} {\bf 72,} 327--346.

\bibitem[\protect\citeauthoryear{Gelfand, Diggle, Guttorp, and Fuentes}{Gelfand
  et~al.}{2010}]{Gelfand2010}
Gelfand, A.~E., Diggle, P.~J., Guttorp, P., and Fuentes, M. (2010).
\newblock {\em {Handbook of spatial statistics}}.
\newblock CRC Press Taylor \& Francis Group, Boca Raton, Florida, USA.

\bibitem[\protect\citeauthoryear{Goldenberg, Zheng, Fienberg, and
  Airoldi}{Goldenberg et~al.}{2010}]{Goldenberg2010}
Goldenberg, A., Zheng, A.~X., Fienberg, S.~E., and Airoldi, E.~M. (2010).
\newblock {A survey of statistical network models}.
\newblock {\em Foundations and Trends{\textregistered} in Machine Learning}
  {\bf 2,} 129--233.

\bibitem[\protect\citeauthoryear{Hastie and Tibshirani}{Hastie and
  Tibshirani}{1986}]{Hastie1986}
Hastie, T. and Tibshirani, R. (1986).
\newblock {Generalized additive models}.
\newblock {\em Statistical Science} {\bf 1,} 297--310.

\bibitem[\protect\citeauthoryear{Hefley, Broms, Brost, Buderman, Kay, Scharf,
  Tipton, Williams, and Hooten}{Hefley et~al.}{2017}]{Hefley2017}
Hefley, T.~J., Broms, K.~M., Brost, B.~M., Buderman, F.~E., Kay, S.~L., Scharf,
  H.~R., Tipton, J.~R., Williams, P.~J., and Hooten, M.~B. (2017).
\newblock {The basis function approach for modeling autocorrelation in
  ecological data}.
\newblock {\em Ecology} {\bf 98,} 632--646.

\bibitem[\protect\citeauthoryear{Hooten, Larsen, and Wikle}{Hooten
  et~al.}{2003}]{Hooten2003}
Hooten, M.~B., Larsen, D.~R., and Wikle, C.~K. (2003).
\newblock {Predicting the spatial distribution of ground flora on large domains
  using a hierarchical Bayesian model}.
\newblock {\em Landscape Ecology} {\bf 18,} 487--502.

\bibitem[\protect\citeauthoryear{Johnson, Conn, Hooten, Ray, and Pond}{Johnson
  et~al.}{2013}]{Johnson2013}
Johnson, D.~S., Conn, P.~B., Hooten, M.~B., Ray, J.~C., and Pond, B.~A. (2013).
\newblock {Spatial occupancy models for large data sets}.
\newblock {\em Ecology} {\bf 94,} 801--808.

\bibitem[\protect\citeauthoryear{K{\'{e}}ry and Royle}{K{\'{e}}ry and
  Royle}{2015}]{Kery2015}
K{\'{e}}ry, M. and Royle, J.~A. (2015).
\newblock {Modeling static occurrence and species distributions using
  site-occupancy models}.
\newblock In {\em Applied Hierarchical Modeling in Ecology}, chapter~10, pages
  551--629. Academic Press, San Diego, CA, USA.

\bibitem[\protect\citeauthoryear{Khan}{Khan}{2013}]{Khan2013}
Khan, S. (2013).
\newblock {Distribution free estimation of heteroskedastic binary response
  models using probit/logit criterion functions}.
\newblock {\em Journal of Econometrics} {\bf 172,} 168--182.

\bibitem[\protect\citeauthoryear{Koenker}{Koenker}{2005}]{Koenker2005}
Koenker, R. (2005).
\newblock {\em {Quantile Regression}}.
\newblock Cambridge University Press, New York, NY, USA.

\bibitem[\protect\citeauthoryear{Koenker and {Bassett Jr.}}{Koenker and
  {Bassett Jr.}}{1978}]{Koenker1978}
Koenker, R. and {Bassett Jr.}, G.~W. (1978).
\newblock {Regression quantiles}.
\newblock {\em Econometrica} {\bf 46,} 33--50.

\bibitem[\protect\citeauthoryear{Komori, Eguchi, Ikeda, Okamura, Ichinokawa,
  and Nakayama}{Komori et~al.}{2016}]{Komori2016}
Komori, O., Eguchi, S., Ikeda, S., Okamura, H., Ichinokawa, M., and Nakayama,
  S. (2016).
\newblock {An asymmetric logistic regression model for ecological data}.
\newblock {\em Methods in Ecology and Evolution} {\bf 7,} 249--260.

\bibitem[\protect\citeauthoryear{Kordas}{Kordas}{2006}]{Kordas2006}
Kordas, G. (2006).
\newblock {Smoothed binary regression quantiles}.
\newblock {\em Journal of Applied Econometrics} {\bf 21,} 387--407.

\bibitem[\protect\citeauthoryear{Kottas and Krnjaji{\'{c}}}{Kottas and
  Krnjaji{\'{c}}}{2009}]{Kottas2009}
Kottas, A. and Krnjaji{\'{c}}, M. (2009).
\newblock {Bayesian semiparametric modelling in quantile regression}.
\newblock {\em Scandinavian Journal of Statistics} {\bf 36,} 297--319.

\bibitem[\protect\citeauthoryear{Larsen, Petersen, Budtz-J{\o}rgensen, and
  Endahl}{Larsen et~al.}{2000}]{Larsen2000}
Larsen, K., Petersen, J.~H., Budtz-J{\o}rgensen, E., and Endahl, L. (2000).
\newblock {Interpreting parameters in the logistic regression model with random
  effects}.
\newblock {\em Biometrics} {\bf 56,} 909--914.

\bibitem[\protect\citeauthoryear{Lee and Neocleous}{Lee and
  Neocleous}{2010}]{Lee2010}
Lee, D. and Neocleous, T. (2010).
\newblock {Bayesian quantile regression for count data with application to
  environmental epidemiology}.
\newblock {\em Journal of the Royal Statistical Society: Series C (Applied
  Statistics)} {\bf 59,} 905--920.

\bibitem[\protect\citeauthoryear{Lemonte and Baz{\'{a}}n}{Lemonte and
  Baz{\'{a}}n}{2018}]{Lemonte2018}
Lemonte, A.~J. and Baz{\'{a}}n, J.~L. (2018).
\newblock {New links for binary regression: An application to coca cultivation
  in Peru}.
\newblock {\em TEST} {\bf 27,} 597--617.

\bibitem[\protect\citeauthoryear{Maalouf and Trafalis}{Maalouf and
  Trafalis}{2011}]{Maalouf2011}
Maalouf, M. and Trafalis, T.~B. (2011).
\newblock {Robust weighted kernel logistic regression in imbalanced and rare
  events data}.
\newblock {\em Computational Statistics and Data Analysis} {\bf 55,} 168--183.

\bibitem[\protect\citeauthoryear{Machado and {Santos Silva}}{Machado and
  {Santos Silva}}{2005}]{Machado2005}
Machado, J.~A. and {Santos Silva}, J.~M. (2005).
\newblock {Quantiles for counts}.
\newblock {\em Journal of the American Statistical Association} {\bf 100,}
  1226--1237.

\bibitem[\protect\citeauthoryear{MacKenzie, Nichols, Hines, Knutson, and
  Franklin}{MacKenzie et~al.}{2003}]{MacKenzie2003}
MacKenzie, D.~I., Nichols, J.~D., Hines, J.~E., Knutson, M.~G., and Franklin,
  A.~B. (2003).
\newblock {Estimating site occupancy, colonization, and local extinction when a
  species is detected imperfectly}.
\newblock {\em Ecology} {\bf 84,} 2200--2207.

\bibitem[\protect\citeauthoryear{MacKenzie, Nichols, Lachman, Droege, Royle,
  and Langtimm}{MacKenzie et~al.}{2002}]{Mackenzie2002}
MacKenzie, D.~I., Nichols, J.~D., Lachman, G.~B., Droege, S., Royle, J.~A., and
  Langtimm, C.~A. (2002).
\newblock {Estimating site occupancy rates when detection probabilities are
  less than one}.
\newblock {\em Ecology} {\bf 83,} 2248--2255.

\bibitem[\protect\citeauthoryear{MacKenzie, Nichols, Royle, Pollock, Bailey,
  and Hines}{MacKenzie et~al.}{2018}]{MacKenzie2018a}
MacKenzie, D.~I., Nichols, J.~D., Royle, J.~A., Pollock, K.~H., Bailey, L.~L.,
  and Hines, J.~E. (2018).
\newblock {\em {Occupancy Estimation and Modeling}}.
\newblock Elsevier.

\bibitem[\protect\citeauthoryear{Manski}{Manski}{1985}]{Manski1985}
Manski, C.~F. (1985).
\newblock {Semiparametric analysis of discrete response: Asymptotic properties
  of the maximum score estimator}.
\newblock {\em Journal of Econometrics} {\bf 27,} 313--333.

\bibitem[\protect\citeauthoryear{McCullagh and Nelder}{McCullagh and
  Nelder}{1983}]{McCullagh1983}
McCullagh, P. and Nelder, J.~A. (1983).
\newblock {\em {Generalized Linear Models}}.
\newblock Chapman and Hall, London, UK.

\bibitem[\protect\citeauthoryear{Meyer}{Meyer}{2008}]{Meyer2008}
Meyer, M.~C. (2008).
\newblock {Inference using shape-restricted regression splines}.
\newblock {\em Annals of Applied Statistics} {\bf 2,} 1013--1033.

\bibitem[\protect\citeauthoryear{Padellini and Rue}{Padellini and
  Rue}{2019}]{Padellini2018}
Padellini, T. and Rue, H. (2019).
\newblock {Model-aware quantile regression for discrete data}.
\newblock {\em arXiv} {\bf 1804,} 1--18.

\bibitem[\protect\citeauthoryear{Prentice}{Prentice}{1976}]{Prentice1976}
Prentice, R.~L. (1976).
\newblock {A generalization of the probit and logit methods for dose response
  curves}.
\newblock {\em Biometrics} {\bf 32,} 761--768.

\bibitem[\protect\citeauthoryear{Reich, Fuentes, and Dunson}{Reich
  et~al.}{2011}]{Reich2011}
Reich, B.~J., Fuentes, M., and Dunson, D.~B. (2011).
\newblock {Bayesian spatial quantile regression}.
\newblock {\em Journal of the American Statistical Association} {\bf 106,}
  6--20.

\bibitem[\protect\citeauthoryear{Royle and K{\'{e}}ry}{Royle and
  K{\'{e}}ry}{2007}]{Royle2007}
Royle, J.~A. and K{\'{e}}ry, M. (2007).
\newblock {A Bayesian state-space formulation of dynamic occupancy models.}
\newblock {\em Ecology} {\bf 88,} 1813--1823.

\bibitem[\protect\citeauthoryear{Royle and Nichols}{Royle and
  Nichols}{2003}]{Royle2003}
Royle, J.~A. and Nichols, J.~D. (2003).
\newblock {Estimating abundance from repeated presence-absence data or point
  counts}.
\newblock {\em Ecology} {\bf 84,} 777--790.

\bibitem[\protect\citeauthoryear{Rue and Held}{Rue and Held}{2005}]{Rue2005a}
Rue, H. and Held, L. (2005).
\newblock {Continuous time random walks}.
\newblock In {\em Gaussian Markov Random Fields: Theory and Application},
  chapter 3.5, pages 123--130. Chapman and Hall/CRC.

\bibitem[\protect\citeauthoryear{Schmid, Zbinden, and Keller}{Schmid
  et~al.}{2004}]{Schmid2004}
Schmid, H., Zbinden, N., and Keller, V. (2004).
\newblock {\em {{\"{U}}berwachung der Bestandsentwicklung h{\"{a}}ufiger
  Brutv{\"{o}}gel in der Schweiz}}.
\newblock Schweizerische Vogelwarte, Sempach.

\bibitem[\protect\citeauthoryear{Schmidt, Rodr{\'{i}}guez, and
  Capistrano}{Schmidt et~al.}{2015}]{Schmidt2015}
Schmidt, A.~M., Rodr{\'{i}}guez, M.~A., and Capistrano, E.~D. (2015).
\newblock {Population counts along elliptical habitat contours: Hierarchical
  modelling using Poisson-lognormal mixtures with nonstationary spatial
  structure}.
\newblock {\em Annals of Applied Statistics} {\bf 9,} 1372--1393.

\bibitem[\protect\citeauthoryear{Sriram, Ramamoorthi, and Ghosh}{Sriram
  et~al.}{2013}]{Sriram2013}
Sriram, K., Ramamoorthi, R.~V., and Ghosh, P. (2013).
\newblock {Posterior consistency of Bayesian quantile regression based on the
  misspecied asymmetric Laplace density}.
\newblock {\em Bayesian Analysis} {\bf 8,} 479--504.

\bibitem[\protect\citeauthoryear{Taddy and Kottas}{Taddy and
  Kottas}{2010}]{Taddy2010}
Taddy, M.~A. and Kottas, A. (2010).
\newblock {A Bayesian nonparametric approach to inference for quantile
  regression}.
\newblock {\em Journal of Business and Economic Statistics} {\bf 28,} 357--369.

\bibitem[\protect\citeauthoryear{Tokdar and Kadaney}{Tokdar and
  Kadaney}{2012}]{Tokdar2012}
Tokdar, S.~T. and Kadaney, J.~B. (2012).
\newblock {Simultaneous linear quantile regression: A semiparametric bayesian
  approach}.
\newblock {\em Bayesian Analysis} {\bf 7,} 51--72.

\bibitem[\protect\citeauthoryear{{Ver Hoef} and Boveng}{{Ver Hoef} and
  Boveng}{2015}]{VerHoef2015}
{Ver Hoef}, J.~M. and Boveng, P.~L. (2015).
\newblock {Iterating on a single sodel is a viable alternative to multimodel
  inference}.
\newblock {\em Journal of Wildlife Management} {\bf 79,} 719--729.

\bibitem[\protect\citeauthoryear{{Ver Hoef}, Peterson, Hooten, Hanks, and
  Fortin}{{Ver Hoef} et~al.}{2018}]{VerHoef2018}
{Ver Hoef}, J.~M., Peterson, E.~E., Hooten, M.~B., Hanks, E.~M., and Fortin,
  M.~J. (2018).
\newblock {Spatial autoregressive models for statistical inference from
  ecological data}.
\newblock {\em Ecological Monographs} {\bf 88,} 36--59.

\bibitem[\protect\citeauthoryear{Wang and Dey}{Wang and Dey}{2010}]{Wang2010}
Wang, X. and Dey, D.~K. (2010).
\newblock {Generalized extreme value regression for binary response data: An
  application to B2B electronic payments system adoption}.
\newblock {\em Annals of Applied Statistics} {\bf 4,} 2000--2023.

\bibitem[\protect\citeauthoryear{Williams, Lu, Scharf, and Hooten}{Williams
  et~al.}{2019}]{Williams2019}
Williams, P.~J., Lu, X., Scharf, H.~R., and Hooten, M.~B. (2019).
\newblock {Bayesian quantile regression for ecological data}.
\newblock {\em unpublished} .

\bibitem[\protect\citeauthoryear{Wood}{Wood}{2017}]{Wood2017}
Wood, S.~N. (2017).
\newblock {\em {Generalized Additive Models: An Introduction with R}}.
\newblock Chapman and Hall/CRC, Boca Raton, Florida, USA, 2 edition.

\bibitem[\protect\citeauthoryear{Wright, Irvine, and Higgs}{Wright
  et~al.}{2019}]{Wright2019}
Wright, W.~J., Irvine, K.~M., and Higgs, M.~D. (2019).
\newblock {Identifying occupancy model inadequacies: Can residuals separately
  assess detection and presence?}
\newblock {\em Ecology} {\bf 0,} e02703.

\bibitem[\protect\citeauthoryear{Xing and Qian}{Xing and Qian}{2017}]{Xing2017}
Xing, J.-J. and Qian, X.-Y. (2017).
\newblock {Bayesian expectile regression with asymmetric normal distribution}.
\newblock {\em Communications in Statistics -- Theory and Methods} {\bf 46,}
  4545--4555.

\bibitem[\protect\citeauthoryear{Yang and Tokdar}{Yang and
  Tokdar}{2017}]{Yang2017}
Yang, Y. and Tokdar, S.~T. (2017).
\newblock {Joint estimation of quantile planes over arbitrary predictor
  spaces}.
\newblock {\em Journal of the American Statistical Association} {\bf 112,}
  1107--1120.

\bibitem[\protect\citeauthoryear{Yang, Wang, and He}{Yang
  et~al.}{2016}]{Yang2016}
Yang, Y., Wang, H.~J., and He, X. (2016).
\newblock {Posterior inference in Bayesian quantile regression with asymmetric
  Laplace likelihood}.
\newblock {\em International Statistical Review} {\bf 84,} 327--344.

\end{thebibliography}

\end{document}